\documentclass[ejsv2,noshowframe]{imsart}
\RequirePackage{amsthm,amsmath,amsfonts}
\RequirePackage[numbers]{natbib}
\usepackage{comment}
\usepackage{dirtytalk}

\RequirePackage[colorlinks,citecolor=blue,urlcolor=blue]{hyperref}
\RequirePackage{graphicx}

\arxiv{2302.09526}
\startlocaldefs
\theoremstyle{plain}

\newtheorem{theorem}{Theorem}[section]
\newtheorem{prop}{Proposition}[section]

\theoremstyle{definition}

\newtheorem{Assump}{Assumption}

\theoremstyle{remark}

\newcommand{\appropto}{\mathrel{\vcenter{
  \offinterlineskip\halign{\hfil$##$\cr
    \propto\cr\noalign{\kern2pt}\sim\cr\noalign{\kern-2pt}}}}}
    
\endlocaldefs

\begin{document}
\begin{frontmatter}
\title{Mixed Semi-Supervised Generalized Linear Regression with Applications to Deep Learning and Interpolators}
\runtitle{Mixed Semi-Supervised Regression}

\begin{aug}
\author[A]{\fnms{Oren}~\snm{Yuval}\ead[label=e1]{orenyuval1@gmail.com}\orcid{0000-0002-3836-4403}},
\author[A]{\fnms{Saharon}~\snm{Rosset}\ead[label=e2]{saharon@tauex.tau.ac.il}}

\address[A]{Department of Statistics and Operation Research,
Tel-Aviv university\printead[presep={,\ }]{e1,e2}}

\runauthor{Yuval and Rosset}
\end{aug}

\begin{abstract}
We present a methodology for using unlabeled data to design semi-supervised learning (SSL) methods that improve the predictive performance of supervised learning for regression tasks. The main idea is to design different mechanisms for integrating the unlabeled data, and include in each of them a \textit{mixing} parameter $\alpha$, controlling the weight given to the unlabeled data. Focusing on Generalized Linear Models (GLM) and linear interpolators classes of models, we analyze the characteristics of different mixing mechanisms, and prove that it is consistently beneficial to integrate the unlabeled data with some nonzero mixing ratio $\alpha>0$, in terms of predictive performance. Moreover, we provide a rigorous framework to estimate the best mixing ratio where mixed-SSL delivers the best predictive performance, while using the labeled and unlabeled data on hand. The effectiveness of our methodology in delivering substantial improvement compared to the standard supervised models, in a variety of settings, is demonstrated empirically through extensive simulation, providing empirical support for our theoretical analysis. We also demonstrate the applicability of our methodology (with some heuristic modifications) to improve more complex models, such as deep neural networks, in real-world regression tasks
\end{abstract}

\begin{keyword}[class=MSC]
\kwd[Primary ]{62F10}
\kwd{62J12}
\kwd[; secondary ]{68T07}
\end{keyword}

\begin{keyword}
\kwd{Predictive modeling}
\kwd{Semi-supervised regression}
\kwd{Generalized linear model}
\kwd{Deep learning}
\end{keyword}

\end{frontmatter}

\section{Introduction}\label{IntroSec}
It is typically more difficult to collect labeled data for training a supervised learning model than to obtain unlabeled data. Hence, there is substantial potential in leveraging unlabeled data to enhance the predictive accuracy of supervised learning through the implementation of suitable semi-supervised learning (SSL) strategies. Over the years, various SSL methods and their effectiveness in improving predictive accuracy have been described (for example \citep{singh2009unlabeled,zhou2014semi,zhu2005semi}), primarily for classification tasks. Recent studies have shown promising results in improving deep neural networks (DNNs) by applying semi-supervised deep learning (SSDL) methods, especially in the field of image classification \citep{sohn2020fixmatch,sun2020deep,han2020unsupervised}. These methods often rely on the clustering assumption (i.e., low data density near decision boundaries, as discussed in \cite{chapelle2005semi}), which is generally not applicable to regression problems. Some works in the literature \citep{wasserman2007statistical,brouard2011semi,niyogi2013manifold,xu2022semi} have explored the efficacy of utilizing unlabeled data to improve metric-based regression models, such as kernel methods. These approaches take advantage of semi-supervised smoothness and manifold assumption (as discussed in \cite{wasserman2007statistical}), which are less relevant to the parametric models that are the main focus of this paper. 

Compared to classification and kernel methods for regression, limited attention has been paid to semi-supervised parametric regression models. In this context, one line of work \citep{tarpey2014paradoxical,chakrabortty2018efficient, zhang2019semi,azriel2022semi} explored how unlabeled data can improve ordinary least squares (OLS) regression. In keeping with statistical tradition, these works are primarily focused on inference regarding the linear coefficients, rather than predictive performance. Another related line of work \citep{javanmard2014confidence,javanmard2018debiasing,bellec2022biasing,bellec2018prediction} discusses SSL methods in Lasso-regularized sparse linear regression. Taking a broad view of these classes of related works, their results suggest that incorporating unlabeled data enhances a linear model’s robustness to misspecification and noise, but predictive performance improvement is not guaranteed in general. Recent works propose SSDL methods for regression \citep{jean2018semi,rezagholiradeh2018reg,olmschenk2019generalizing,he2022semi}, incorporating an 'unsupervised loss' term derived from unlabeled data. In particular, \cite{jean2018semi} suggests using the predictive variance of the unlabeled data as a regularization term, added to the objective during training, scaled by a regularization parameter. Although the method's intuition is well explained and linked to \textit{regularized Bayesian inference} (\cite{zhu2014bayesian}), theoretical analysis is challenging and does not definitively show superiority over supervised models. Instead, the authors empirically demonstrate the benefits of SSDL through systematic experiments on real-world datasets.

In this paper, we suggest adding an unsupervised loss to the loss function of generalized linear models (GLM), thereby creating a general form of semi-supervised objective function. Furthermore, and more importantly, we quantify the extent to which unlabeled data are incorporated through a \textit{mixing ratio} $\alpha$, thereby introducing the concept of \textit{mixed-SSL}. In this framework, we are able to characterize the performance of the supervised and semi-supervised estimators, and obtain theoretical insights. Our main result states that it is invariably beneficial to integrate the unlabeled data with some nonzero mixing ratio $\alpha>0$, in terms of predictive performance. Moreover, we provide a constructive methodology for estimating the optimal mixing ratio, where mixed-SSL delivers the best predictive performance while using the data on hand, and demonstrate our methodology empirically through extensive simulations. This result differs from traditional studies on SSL in linear models, which suggest that unlabeled data do not always reduce the predictive error \citep{tarpey2014paradoxical,chakrabortty2018efficient,azriel2022semi}. In addition, the findings of this study provide an explanation for the observed benefit of SSDL in regression tasks, typically involving a GLM as the output layer of a deep neural network. We also demonstrate the applicability of our methodology to improve state-of-the-art supervised deep learning models in real-world regression tasks.

The methodology and theory presented in this paper improve and expand our previous work (\cite{yuval2022semi}), which proposed a general methodology for semi-supervised empirical risk minimization. A major contribution of that work was a characterization of the relative performance of the supervised and semi-supervised estimators, and a consequent methodology for choosing between them, according to appropriate data-based estimates of the noise and signal of the model. In this paper, we show that mixing between the estimators leads to better performance than selecting between them. Furthermore, we broaden the discussion to include interpolator models, offering theoretical and empirical evidence on the advantages of the proposed mixed-SSL approach in this case as well.

\subsection{Setup and notations}
In the supervised framework, a training sample is given by $T=(X,Y)$, where $X\in \mathbb{R}^{n\times p}$ with rows $x_1,\ldots\ ,x_n$, and $Y=(y_1,\ldots\ ,y_n)^t \in \mathbb{R}^n$ such that the pairs $(x_1,y_1),\ldots\,(x_n,y_n)$ are i.i.d. according to some joint distribution $P_{xy}$. In the semi-supervised setting, the training sample is given by $T=(X,Y,Z)$, where the additional matrix $Z\in \mathbb{R}^{m\times p}$ represents the set of unlabeled data with $z_1,\ldots\,z_m$ i.i.d.\ observations drawn from the marginal distribution $P_{x}$.
Initially, we examine the scenario when $p<n$, and subsequently in Section \ref{Interpolators-Sec-New}, we analyze interpolating models for the case where $p>n$. For simplicity, we assume that the distribution $P_{x}$ is centered around zero, that is, $\mathbb{E}[x] = \textbf{0}_p$ and that $\mathbb{V}ar(x)= \Sigma$ is a positive definite matrix. We also assume that the distribution $P_{y \mid x}$ follows a GLM \cite{nelder1972generalized}:
\[ \mathbb{E}[y \mid x] = g(x^t\beta) \hspace{2mm};\hspace{2mm}  \mathbb{V}ar(y \mid x) = g'(x^t\beta)\sigma^2, \]
for some $\beta \in \mathbb{R}^p$, where $g$ is a known, strictly increasing \textit{link function}, and $\sigma^2$ refers to the \say{noise} of the underlying model of the data. For a given training set $T$, we are interested in defining a statistical learning process that fits an estimator $\beta_T$ of $\beta$.

Consider an independent sample $(x_0,y_0)$ from the joint distribution $P_{xy}$, and a linear predictor $\beta_T$ that is learned based on the training set $T$. In the context of GLM, the objective is to minimize the out-of-sample predictive error (this corresponds to maximizing the log likelihood of $\beta$ given $(x_0,y_0)$,  as described in detail in \cite{nelder1972generalized}): 
\begin{align} \label{Rdef0}
    R_{\beta}(\beta_T) &= \mathbb{E}_{T,x_0,y_0}\left[G(x_0^t\beta_T) - x_0^t\beta_T y_0 \right] \\ \label{Rdef1} & = \mathbb{E}_{T,x_0}\left[G(x_0^t\beta_T) - x_0^t\beta_T g(x_0^t\beta) \right],
\end{align}
where $G'=g$. Assuming interchangeability of derivative and expectation, which we adopt throughout this paper, it is straightforward to demonstrate that $R_{\beta}(\beta_T)$ is minimized at the unique point $\beta_T = \beta$. 

As elaborated in Appendix~\ref{appC}, by applying a quadratic approximation to the convex loss function $R_{\beta}(\beta_T)$ around the true parameter vector $\beta$, the predictive error can be written in the following approximate formula:
\begin{align}\label{r_beta_def}
    R_{\beta}(\beta_T)&\approx R_{\beta}(\beta) + \frac{1}{2n}\mathbb{E}_{T}\left[ (\beta_T-\beta)^tH(\beta_T-\beta)\right]:= R_{\beta}(\beta)+r_{\beta}(\beta_T),
\end{align}
where $R_{\beta}(\beta)$ is a constant (irreducible) term and $r_{\beta}(\beta_T)$ is the \textit{reducible error} that distinguishes between different estimators, $H= \mathbb{E}_X[X^tDX]$, and $D$ is a diagonal $n\times n$ random matrix with elements $D_{ii}=g'(x_i^t\beta)$. The notation $\approx$ denotes the imprecision introduced by the quadratic approximation, which arises by neglecting higher-order terms. This imprecision depends on the magnitude of the derivatives of $g$ beyond the first order, as further discussed in Appendix~\ref{appC}. Similar approximation methods for nonlinear statistical models appear early in the literature \cite{efron1978assessing}, and later in \cite{efron2018curvature}. In the special case of a linear model (LM), where $g$ serves as the identity function, the function $R_{\beta}(\beta_T)$ is quadratic, thus eliminating any imprecision in the given formulation. Assuming the quadratic error in Equation (\ref{r_beta_def}) is negligible, it is therefore natural to examine the characteristics of $r_{\beta}(\beta_T)$ to understand the impact of different SSL methods on predictive performance.

The motivation for SSL arises when the amount of unlabeled data greatly exceeds the amount of labeled data, that is, when $m \gg n,p$. In such cases, it is reasonable to assume that the expectation $\mathbb{E}_x\left[\varphi (x) \right]$ of a function of interest $\varphi$ can be well evaluated by its empirical average over the unlabeled set $Z$, i.e., $\frac{1}{m} \sum_{i=1}^m \varphi (z_i)$. This motivates what we refer to as the \textit{total information} assumption, meaning that the expectations $\mathbb{E}_x\left[\varphi (x) \right]$ of all relevant functions $\varphi$ are known with high accuracy. We adopt this assumption in our theoretical analysis primarily for clarity and analytical tractability. However, Appendix \ref{appB} presents a brief analysis without this assumption, showing that our core theoretical conclusions still hold as long as the amount of unlabeled data is sufficiently large. Furthermore, in our empirical simulations, we rely on large but finite unlabeled datasets, and the results indicate that relaxing the total information assumption does not substantially impact practical performance.

\subsection{Outline of the paper}
In section \ref{semi-supervised-GLM}, we introduce the mixed-SSL method to enhance GLM for scenarios where $p<n$. This involves the mixing parameter $\alpha$, which indicates the degree to which the unlabeled data are incorporated into the learning procedure. Our approximate analysis indicates that, in general, integrating unlabeled data into the learning process (i.e., choosing $\alpha>0$) is invariably advantageous, with the benefits amplifying as the noise level $\sigma^2$ grows. In particular, for LM with $\mathbb{E}[y \mid x]= x^t\beta$ this statement holds precisely. This theoretical result stands in marked contrast to earlier studies \citep{yuval2022semi,azriel2022semi,chakrabortty2018efficient} which stated that the unlabeled data are useful only when the model is biased, nonlinear, or extremely noisy. We further evaluate the effectiveness of unlabeled data in interpolating models where $p>n$ (Section \ref{Interpolators-Sec-New}). This is done by presenting the \textit{mixed-SSL interpolator} and providing theoretical evidence of the superiority of our approach over the standard minimum-norm interpolator.

In Section \ref{Estimating-Methods}, we propose methods for tuning the mixing parameter for each of the settings in Sections \ref{semi-supervised-GLM},~\ref{Interpolators-Sec-New}  using the available data, leading to data-driven, adaptive mixed-SSL estimators that are practical for use. In Section \ref{Assimp-New}, we examine the attributes of the proposed mixed-SSL methods in an asymptotic setup
and show that the advantage of utilizing the unlabeled data persists as the amount of labeled data diverges. All the theoretical results described above are supported by empirical results from experiments carried out with simulated data (Section~\ref{Results}). Additionally, we observe that our method for choosing an appropriate mixing ratio is successful in enhancing the prediction performance. In Section~\ref{Results-Real}, we present intuitive modifications to our framework to tackle complex models like DNNs and empirically demonstrate the benefits of using mixed-SSDL to enhance performance on real-world datasets.

\section{Mixed semi-supervised GLM}\label{semi-supervised-GLM}
In this section, we focus on the GLM framework for regression tasks in the setting $p < n$. In this case, the standard estimator for $\beta$ is obtained via empirical risk minimization of the predictive error (\ref{Rdef0}) on the labeled data, which can be written as follows:
\begin{align} \label{L_hat_Def}
    \hat{L}(\beta;X,Y) = \frac{1}{n}\sum_{i=1}^n G(x_i^t\beta)-x_i^t\beta y_i,
\end{align}
leading to the following optimization problem:
\begin{align}\label{e1}
\hat{\beta} &= \underset{\beta\in \mathbb{R}^p}{\arg \min} \left\{ \hat{L}(\beta;X,Y) \right\}. 
\end{align} 
Building on the total information assumption, \cite{yuval2022semi} proposed a semi-supervised loss function, which incorporates information from the distribution $P_x$ as follows:
\begin{align} \label{L_breve_Def}
     \Breve{L}(\beta;X,Y)=  \mathbb{E}_x\left[G(x^t\beta)\right]- \mathbb{E}_{x}  \left[ x^t\beta\right] \overline{Y} - \widehat{\mathbb{C}ov}(X\beta,Y),
\end{align}
where,
\begin{align*}
    \widehat{\mathbb{C}ov}(X\beta,Y)  = \sum_{i=1}^n\left( x_i^t\beta-  \overline{X\beta} \right) \left(  y_i - \overline{Y} \right)/n \hspace{1mm};\hspace{1mm}
    \overline{Y}=\frac{1}{n}\sum_{i=1}^n y_i \hspace{1mm};\hspace{1mm} \overline{X\beta} =\frac{1}{n} \sum_{i=1}^n x_i^t\beta,
\end{align*}
and the semi-supervised estimator is defined as follows:
\begin{align}\label{e3}
     \Breve{\beta} &= \underset{\beta\in \mathbb{R}^p}{\arg\min} \left\{  \Breve{L}(\beta;X,Y) \right\} 
\end{align}

Analysis in \cite{yuval2022semi} showed that $\breve{\beta}$ outperforms $\hat{\beta}$ only when the noise level $\sigma^2$ exceeds a specific threshold. This motivated the adaptive estimator $\beta^D$ which selects $\breve{\beta}$ when the estimated noise variance is high and otherwise defaults to $\hat{\beta}$. Although empirical results suggest that $\beta^D$ reduces predictive error, its theoretical justification remains unclear.

Now, we move beyond a binary choice between $\hat{\beta}$ and $\breve{\beta}$, and propose two mixed-SSL parameterized by $\alpha$. The first one is the linear-mixed estimator, which is essentially a linear mixing between the supervised and semi-supervised estimators: 
\begin{align}\label{e5}
\dot{\beta}_{\alpha}=(1-\alpha)\hat{\beta}+\alpha \breve{\beta},
\end{align}
where $\alpha \in [0,1]$. The second procedure mixes the two loss functions to form:
\begin{align} \nonumber 
     L^M_{\alpha}(\beta,X,Y) &= (1-\alpha)\hat{L}(\beta,X,Y)+\alpha \breve{L}(\beta,X,Y) \\  \label{L_Mixed_Def} &=  \frac{1}{n}\sum_{i=1}^n (1-\alpha)G(x_i^t\beta)-x_i^t\beta y_i  +\alpha \left(\mathbb{E}[G(x^t\beta)] + \overline{X\beta}\cdot \overline{Y} \right),
\end{align}
leading to the following optimization problem:
\begin{align} 
     \nonumber \Ddot{\beta}_{\alpha} = \underset{\beta\in \mathbb{R}^p}{\text{{\footnotesize argmin}}} \Big\{  L^M_{\alpha}(\beta,X,Y) \Big\}.  
\end{align}

In both learning procedures, the mixing ratio $\alpha$ quantifies the degree of integration of unlabeled data in training. The mixed loss function, $L^M_{\alpha}$, is equivalent to $\hat{L}+\frac{\alpha}{1-\alpha}\breve{L}$, for $\alpha < 1$, thus it can be viewed as a regularized supervised learning with the regularization term determined by the unlabeled data. This idea is related to the concept of penalizing predictive variance, which was introduced in \cite{jean2018semi}. However, the regularization term suggested here is more general in the sense that it can be applied to different link functions $g$. In fact, the predictive variance penalty is a special case of the first penalty term in $L^M_{\alpha}$, where $g$ is the identity, and therefore: $\mathbb{E}_x\left[G(x^t\beta)\right]=0.5\mathbb{V}ar_x\left(f(x) \right)$. 

In this section, we aim to analyze the predictive performance of the proposed estimators, $\Ddot{\beta} _{\alpha}$ and $\dot{\beta} _{\alpha}$, compared to that of $\hat{\beta}$. We introduce the following bias-variance decomposition of the reducible error for an arbitrary estimator $\beta_T$:
\begin{align*}
    r_{\beta}(\beta_T)=& \underbrace{ \frac{1} {2n}\text{tr}\left( H \mathbb{E}_{X}\left[ \left(\mathbb{E}[\beta_T \mid X] - \beta   \right)\left(\mathbb{E}[\beta_T \mid X] - \beta   \right)^t \right]  \right) }_\text{$B_{\beta}(\beta_T)$} + \underbrace{  \frac{1} {2n}\text{tr}\left( H  \mathbb{E}_{X}\left[ \mathbb{V}ar\left( \beta_T \mid X\right)\right] \right)
  }_\text{$V_{\beta}(\beta_T)$}.
\end{align*}
In the above decomposition, the expression represented by $B_{\beta}(\beta_T)$ signifies the bias of $\beta_T$ compared to $\beta$, while $V_{\beta}(\beta_T)$ signifies the variance of the estimator. The following assumption ensures that $V_{\beta}(\hat{\beta})$ is well defined and finite, and will be used throughout our analysis.
\begin{Assump}\label{A0n} 
For any fixed vector $\beta$, the distribution $P_x$ and the link function $g$ satisfy that $(X^tDX)^{-1}$ exists with probability $1$, and also that $\mathbb{E}_{X}\left[ (X^tDX)^{-1}\right]$ exists.
\end{Assump}

Starting with our examination of $\dot{\beta}_{\alpha}$, we are looking for an approximate closed form for its bias and variance. As elaborated in Appendix~\ref{appC}, by using the quadratic approximation of the convex loss functions $\hat{L}(\beta;X,Y)$ and $\Breve{L}(\beta;X,Y)$ around the true vector $\beta$, we find that:
\begin{align} \label{dot_B_Def}
    B_{\beta}(\dot{\beta}_{\alpha}) &\approx \dot{B}_{\beta}(\alpha) = \frac{\alpha^2}{2n} \text{tr}\left(H^{-1}  \mathbb{E}_{X}\left[ \zeta \zeta^t \right]   \right),
    \\  \label{dot_V_Def}  V_{\beta}(\dot{\beta}_{\alpha}) &\approx  \dot{V}_{\beta}(\alpha)  = \frac{\sigma^2}{2} \left(\alpha^2 v_u +(1-\alpha)^2v_l +2\alpha(1-\alpha) v_u  \right),
\end{align}
where
\begin{align*}
    \zeta=\mathbb{E}_X[X^t\mu] -n\widehat{\mathbb{C}ov}(X,\mu) \hspace{1mm};\hspace{1mm} v_u= \frac{n-1}{n^2} p \hspace{1mm};\hspace{1mm} v_l=\frac{1}{n}\text{tr}\left( \mathbb{E}_X\left[ (X^tDX)^{-1}  \right]H \right),
\end{align*}
    where $\mu \in \mathbb{R}^n$ has elements $\mu_{i}=g(x_i^t\beta)$, and $\widehat{\mathbb{C}ov}(X,\mu)$ is a random vector in $\mathbb{R}^p$, with the component $\widehat{\mathbb{C}ov}\left( [X]_j, \mu \right)$ in the $j^\text{th}$ entry, where $[X]_j$ represents the $j^\text{th}$ column of $X$. Here, the notation $\approx$ denotes the imprecision introduced by neglecting the terms of order higher than two in $\hat{L}(\beta;X,Y)$ and $\Breve{L}(\beta;X,Y)$.

We denote $\dot{r}_{\beta}(\alpha) = \dot{B}_{\beta}(\alpha) + \dot{V}_{\beta}(\alpha)$ as the function that roughly represents the relationship between $\alpha$ and the predictive performance of $\dot{\beta}_{\alpha}$. We note that for the LM there are no approximations involved and analyzing $\dot{r}_{\beta}(\alpha)$ is equivalent to analyzing $R_{\beta}(\dot{\beta}_{\alpha})$ (they differ only by the irreducible error). Thus, we aim to understand how $\dot{r}_{\beta}(\alpha)$ depends on $\alpha$. The following theorem shows that the reducible error cannot be minimized at the boundary values of $\alpha=0$ or $\alpha=1$.
\begin{theorem}\label{P2n} Under Assumption~\ref{A0n}, the unique global minimizer of $\dot{r}_{\beta}(\alpha)$, denoted by $\dot{\alpha}$, admits the following explicit formula:
\begin{align}\label{alp_dot_formula}
    \dot{\alpha}=\frac{\sigma^2(v_l-v_u) }{\dot{B}_{\beta}(\alpha=1)+\sigma^2(v_l-v_u)},
\end{align}
and the equality $v_l-v_u >0$ holds, ensuring that $\dot{\alpha}$ is in the open interval $(0,1)$, for any instance of $(\beta, \sigma^2)$. Moreover, the minimum value of $\dot{r}_{\beta}(\alpha)$ denoted by $\dot{r}_{\beta}(\dot{\alpha})$, admits the following explicit formula:
\begin{align*}
    \dot{r}_{\beta}(\dot{\alpha})= \frac{1}{2} \left[\sigma^2v_l -\frac{\sigma^4(v_l-v_u)^2 }{\dot{B}_{\beta}(\alpha=1)+\sigma^2(v_l-v_u)} \right]< \dot{r}_{\beta}(\alpha=0).
\end{align*}
\end{theorem}

The proof of Theorem~\ref{P2n} is given in Appendix \ref{appA1}, where it also reveals a bias-variance tradeoff with respect to the mixing parameter $\alpha$. As $\alpha$ increases, the approximate bias $\dot{B}_{\beta}(\alpha)$ increases, while the approximate variance $\dot{V}_{\beta}(\alpha)$ decreases, due to the fact that $v_u<v_l$. The optimal value $\dot{\alpha}$ balances these opposing effects, identifying the ideal level of integration of unlabeled data in the learning process. Notably, the fact that $\dot{\alpha}$ increases with the noise level aligns with previous findings, suggesting that the benefits of incorporating unlabeled data are most pronounced in high-noise settings. In conclusion, coupled with the lack of approximations for the LM, Theorem~\ref{P2n} fully demonstrates the advantage of $\dot{\beta}$ over the supervised estimator $\hat{\beta}$ within the framework of LM.

We shift our attention to the characteristics of $\ddot{\beta}_{\alpha}$ and also consult Appendix~\ref{appC}. Analogous to $\dot{\beta}_{\alpha}$, employing a quadratic approximation of $L^M_{\alpha}$ around $\beta$ provides approximate closed forms for the bias and variance of $\ddot{\beta}_{\alpha}$.
\begin{align} \label{ddot_B_Def}
    B_{\beta}(\ddot{\beta}_{\alpha}) &\approx \ddot{B}_{\beta}(\alpha) = \frac{\alpha^2}{2n}\text{tr}\left(\mathbb{E}_{X}\left[ S_{\alpha}  H S_{\alpha} \zeta \zeta^t \right]\right),
    \\  \label{ddot_V_Def}  V_{\beta}(\ddot{\beta}_{\alpha}) &\approx  \ddot{V}_{\beta}
    (\alpha)  = \frac{ \xi_{\alpha}\sigma^2}{2n}\text{tr}\left(\mathbb{E}_X\left[ S_{\alpha}HS_{\alpha} X^tDX\right] \right),
\end{align}
where 
\[S_{\alpha}=  \left(\alpha  H+ (1-\alpha)X^tDX\right  )^{-1} \hspace{2mm};\hspace{2mm} \xi_{\alpha}=1-(2\alpha-\alpha^2)/n.\]
We denote $\ddot{r}_{\beta}(\alpha) = \ddot{B}_{\beta}(\alpha) + \ddot{V}_{\beta}(\alpha)$ as the function that roughly represents the relationship between $\alpha$ and the predictive performance of $\ddot{\beta}_{\alpha}$. Similarly to $\dot{r}_{\beta}(\alpha)$, for the LM there are no approximations involved, and analyzing $\ddot{r}_{\beta}(\alpha)$ is equivalent to analyzing $R_{\beta}(\ddot{\beta}_{\alpha})$. Therefore, our objective is to comprehend the behavior of $\ddot{r}_{\beta}(\alpha)$ with respect to changes in $\alpha$. The following theorem indicates that the approximate reducible error, $\ddot{r}_{\beta}(\alpha)$, is not optimal at the extremes of $\alpha=0$ or $\alpha=1$, and provides insights into the relationship between $\ddot{V}_{\beta}(\alpha)$ and $\dot{V}_{\beta}(\alpha)$.

\begin{theorem}\label{P4n}
     Under Assumption \ref{A0n}, for any $(\beta,\sigma^2)$, the minimizer of $\ddot{r}_{\beta}(\alpha)$, denoted by $\ddot{\alpha}$, is in the open interval $(0,1)$. Moreover, at any given mixing ratio $\alpha$, the following inequality holds: $\ddot{V}_{\beta}(\alpha) \leq \dot{V}_{\beta}(\alpha)$.
\end{theorem}
The proof of Theorem~\ref{P4n} can be found in Appendix \ref{appA4n}. From this theorem, we conclude that $\Ddot{r}_{\beta}(\alpha)$ behaves similarly to $\dot{r}_{\beta}(\alpha)$ but is more robust to noise, since its variance component is uniformly smaller. In the particular case of LM, the simplifications in the bias terms $B_{\beta}(\ddot{\beta}_{\alpha})$ and $B_{\beta}(\dot{\beta}_{\alpha})$ enable us also to gain understanding about the relationship between these two terms. We explore this connection under a wide range of distributions $P_x$, which satisfy the following assumption.
\begin{Assump}\label{AMechanisim_n} 
The distribution $P_x$ satisfies the following generating mechanism: $x=\Sigma^{1/2}\tilde{x}$, where $\tilde{x}\in \mathbb{R}^p$ is a random vector with i.i.d. entries that have zero mean, unit variance, and a finite fourth moment $q_4$, a finite sixth moment $q_6$, and $\Sigma$ is a positive definite covariance matrix.
\end{Assump}

\begin{theorem}\label{P2.11n}
    In the LM, under Assumptions \ref{A0n} and \ref{AMechanisim_n}, for any combination $(\beta, \sigma^2)$, and given mixing ratio $\alpha$, the following inequality holds:
    \[ B_{\beta}(\ddot{\beta}_{\alpha}) \approx \Ddot{b}(\alpha) \leq B_{\beta}(\dot{\beta}_{\alpha}),\]
    where $\Ddot{b}(\alpha)$ is a third-order polynomial approximation of $B_{\beta}(\ddot{\beta}_{\alpha})$ that satisfies the following constraints:
    \begin{align*}
        \Ddot{b}(0)=  B_{\beta}(\ddot{\beta}_{\alpha=0})   \hspace{1mm};\hspace{1mm}  \ddot{b}'(0)= \frac{\partial B_{\beta}(\ddot{\beta}_{\alpha})}{\partial \alpha} \mid _{\alpha=0},   \\
        \Ddot{b}(1)= B_{\beta}(\ddot{\beta}_{\alpha=1}) \hspace{1mm};\hspace{1mm} \ddot{b}'(1) = \frac{\partial B_{\beta}(\ddot{\beta}_{\alpha})}{\partial \alpha}  \mid _{\alpha=1}.
    \end{align*}  
\end{theorem}
The proof of Theorem~\ref{P2.11n} is provided in Appendix \ref{appA4}. Combining also Theorems \ref{P2n} and \ref{P4n}, we deduce that in the LM the following ordering holds for any $\beta$ and $\sigma^2$: 
\begin{align}\label{T_Chain}
    R_{\beta}(\ddot{\beta}_{\ddot{\alpha}}) \leq R_{\beta}(\ddot{\beta}_{\dot{\alpha}}) \lesssim R_{\beta}(\dot{\beta}_{\dot{\alpha}}) \leq R_{\beta}(\hat{\beta}),
\end{align}
where the notation $\lesssim$ denotes an approximate assessment of $B_{\beta}(\ddot{\beta}_{\alpha})$ using a third-order approximation. If the model is a GLM and satisfies the condition that the quadratic error is negligible in relation to the characteristics of $R_{\beta}(\dot{\beta}_{\alpha})$ and $R_{\beta}(\ddot{\beta}_{\alpha})$ with respect to $\alpha$, then it is probable that the sequence of inequalities in (\ref{T_Chain}) will hold true as well.

\begin{figure}
\centering
\includegraphics[width=12cm]{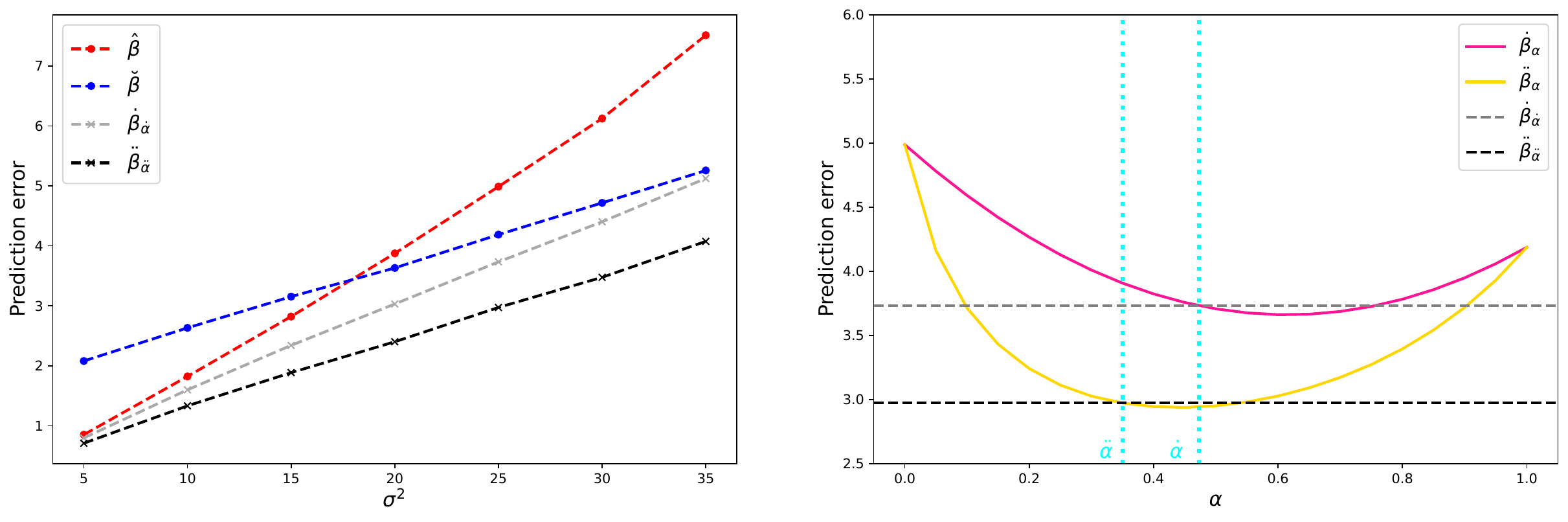}
\caption{Mean predictive errors of supervised and semi-supervised estimators in ELU model with $n=50$ and $p=10$. On the left side, we examine various values of $\sigma^2$. On the right side, $\sigma^2=25$ and we examine various values of $\alpha$ along the $[0,1]$ grid. see text for details.}
\label{MainRes1}
\end{figure}

Figure \ref{MainRes1} displays the primary findings from the empirical study carried out using simulated data, demonstrating the characteristics of the mixed-SSL in a GLM with the ELU link function, i.e., $g(x^t\beta)=\min\left\{e^{x^t\beta}-1,\max\left(0,x^t\beta\right)\right\}$. On the left panel, it is apparent that for any given $\sigma^2$, the chain of inequalities shown in (\ref{T_Chain}) holds. On the right panel, we can see that both $\dot{\alpha}$ and $\Ddot{\alpha}$ serve as excellent approximations for identifying the optimal mixing ratios of $\dot{\beta}_{\alpha}$ and $\Ddot{\beta}_{\alpha}$ respectively. This holds true despite their roles as the minimizers of $\dot{r}_{\beta}(\alpha)$ and $\ddot{r}_{\beta}(\alpha)$, rather than of $R_{\beta}(\dot{\beta}_{\alpha})$ and $R_{\beta}(\ddot{\beta}_{\alpha})$. Moreover, $\dot{\alpha}$ determines an appropriate mixing ratio for $\Ddot{\beta}_{\alpha}$. The complete information regarding the empirical study, along with additional results, is given in Section \ref{GLM_Simulations}.

\section{Mixed semi-supervised interpolators}\label{Interpolators-Sec-New}
In this section, we extend our methodology to the over-parameterized regime, focusing on \textit{interpolators} --- estimators that achieve zero training error, such as DNNs. These models have attracted increasing attention in recent years. This is mainly because state-of-the-art DNNs appear to be models of this type, and because the recently observed double descent phenomenon shows that out-of-sample error can decrease again as the number of parameters grows \citep{belkin2019reconciling,geiger2019jamming,hastie2022surprises}. In the last paper, the authors analyze the minimum $l_2$ norm interpolation in high-dimensional least squares regression and show that double descent can occur in this model under a variety of settings.

The basic principle in this type of models is to use a large number of parameters $p\gg n$, and to interpolate the training data, while using an appropriate criterion to choose the \say{best} interpolator, typically based on norm minimization. 
In practice, deep learning models are fitted by first-order methods, such as gradient descent, which, when initialized appropriately, generally converges to an interpolator with a nearly minimal distance from the initial point (see \cite{oymak2019overparameterized}). Moreover, as shown by \cite{vardi2021implicit}, while the implicit regularization of a ReLU neuron is not expressible, it is approximately the $l_2$ norm. These properties explain why first-order methods yield reliable learning in modern over-parameterized regimes. For our purposes, they suggest that an SSL approach which improves the minimal $l_2$-norm interpolator may also enhance deep learning models.

Our study on over-parameterized mixed-SSL examines the case involving a true linear model, $\mathbb{E}[y \mid x,\beta]=x^t\beta$, in which the vector $\beta$ is treated as a random variable. We assume that for each training sample $T=(X,Y)$, $\beta$ is drawn independently from a prior distribution $P_\beta$ with $\mathbb{E}\left[ \beta \right] = \textbf{0}_p$ and $\mathbb{E}\left[ \beta \beta^t \right] = \tau^2 I_p$, where $\tau^2$ represents the signal strength of the model. In this case, the mean predictive error of an estimator $\beta_T$ is obtained by averaging over $P_\beta$, yielding:
\[R(\beta_T)=\mathbb{E}_{\beta}\left[ R_{\beta}(\beta_T)  \right] =  \mathbb{E}_{\beta}\left[R_{\beta}(\beta)+r_{\beta}(\beta_T)\right]= R(\beta)+r(\beta_T),\]
and by substituting into the linear model, we obtain the subsequent decomposition for the mean reducible error:
\begin{align*}
    r(\beta_T) =& \hspace{1mm} \underbrace{\frac{1}{2} \mathbb{E}_{X,x_0,\beta}\left[ \left( x_0^t\mathbb{E}[\beta_T \mid X,\beta] - x_0^t\beta   \right)^2 \right]   }_\text{$B(\beta_T)$}+ \underbrace{\frac{1}{2}  \mathbb{E}_{X,x_0,\beta}\left[ \mathbb{V}ar\left( x_0^t\beta_T \mid X,x_0,\beta\right)\right] 
  }_\text{$V(\beta_T)$}.
\end{align*}


Analogous to the variance penalty in the LM case, we propose using unlabeled data to fit a linear interpolator that reduces predictive variance. This interpolator is then mixed with the minimum-norm interpolator to achieve an interpolating model with enhanced predictive performance. The standard supervised minimum-norm interpolator takes the following explicit form:
\begin{equation*}
    \hat{\beta}^I = \underset{\beta\in \mathbb{R}^p}{\arg\min} \left\{  || \beta ||_2 \hspace{2mm} : \text{ s.t. } X\beta=Y   \right\} = X^t(XX^t)^{-1}Y. 
\end{equation*}
Assuming that $\mathbb{V}ar(x)=\Sigma$ is known from the unlabeled data, we define the semi-supervised estimator in this context as the \textit{minimum-variance interpolator}, expressed as follows:
\begin{align*}
    \tilde{\beta}^I &= \underset{\beta\in \mathbb{R}^p}{\arg\min} \left\{ \beta^t\Sigma \beta \hspace{2mm}, \text{ s.t. } X\beta=Y   \right\} = \Sigma^{-1}X^t(X\Sigma^{-1}X^t)^{-1}Y. 
\end{align*}
Now, let us denote the \textit{linear-mixed-interpolator} by  \[\dot{\beta}^I_{\alpha} = (1-\alpha)\hat{\beta}^I+\alpha \tilde{\beta}^I,\]
for some mixing ratio $\alpha \in [0,1]$. We next analyze the properties of $\dot{\beta}^I_{\alpha}$ and examine the advantages of incorporating unlabeled data in interpolating models.

To analyze the properties of the mean predictive error of our suggested mixed estimator, $R(\dot{\beta}^I_{\alpha})$, we require the following distributional assumption:  
\begin{Assump}\label{A5} 
The distribution $P_x$ satisfies that $(XX^t)^{-1}$, $(XX^t)^{-2}$, and $(X\Sigma X^t)^{-1}$ exist with probability $1$, and also that their expectations exist.
\end{Assump}

\begin{prop}\label{P5n}
    In the case of the linear interpolating model, under Assumption \ref{A5}, for any $\alpha\in [0,1]$, the reducible error of $\dot{\beta}^I_{\alpha}$ can be written as follows:
    \[ r(\dot{\beta}^I_{\alpha})= \underbrace{\frac{\tau^2}{2} \left[  \alpha^2(b_u^I-b_l^I)+\mathrm{tr}(\Sigma)-b_u\right]  }_\text{$B(\dot{\beta}^I_{\alpha})$} + \underbrace{\frac{\sigma^2}{2} \left[ (1-\alpha)^2v_l^I  + (2\alpha-\alpha^2) v_u^I   \right]
  }_\text{$V(\dot{\beta}^I_{\alpha})$},\]
    where
    \begin{align*}
        b_u^I=\text{tr}\left(  \Sigma \mathbb{E}\left[X^t(XX^t)^{-1}X\right] \right) \hspace{2mm}&;\hspace{2mm} b_l^I=\text{tr}\left( \mathbb{E}\left[X^t(X\Sigma^{-1} X^t)^{-1}X\right]   \right), \\ v_u^I = \text{tr}\left(\mathbb{E}_X\left[ (X\Sigma^{-1}X^t)^{-1} \right]  \right) \hspace{2mm}&;\hspace{2mm} v_l^I =\text{tr}\left(\Sigma  \mathbb{E}_X\left[X^t(XX^t)^{-2}X\right]  \right).
    \end{align*}  

    Moreover, for any distribution $P_x$ satisfying Assumption~\ref{A5}, the following holds: $b^I_l\leq b^I_u$ \textbf{and} $v^I_l\geq v^I_u$. 
\end{prop}
The proof of Proposition~\ref{P5n} is provided in Appendix \ref{appA110}, and from it we deduce the relationship between the mixing ratio $\alpha$ and the predictive error associated with the mixed estimator $\dot{\beta}^I_{\alpha}$. In particular, since $b^I_l\leq b^I_u$ and $v^I_l\geq v^I_u$, Proposition~\ref{P5n} highlights a bias-variance tradeoff concerning $\alpha$. As $\alpha$ increases, the bias term $B(\dot{\beta}^I_{\alpha})$ increases and the variance term $V(\dot{\beta}^I_{\alpha})$ decreases. Thus, as in the previous section, the minimizer of $r(\dot{\beta}^I_{\alpha})$, denoted by $\dot{\alpha}_I$, balances  between bias and variance, identifying the optimal level of unlabeled data integration. We note that the equalities $b^I_l = b^I_u$ and $v^I_l = v^I_u$ hold in the edge case where $\hat{\beta}^I=\tilde{\beta}^I$ for any given realization of $(\beta,X,Y)$, which occurs when $\Sigma=\lambda I_p$, for some $\lambda>0$.

\begin{theorem}\label{Interpol1} For any distribution $P_x$ satisfying Assumption~\ref{A5}, and given parameters $(\sigma^2,\tau^2)$, the unique global minimum of $r(\dot{\beta}^I_{\alpha})$ (and $R( \dot{\beta}^I_{\alpha})$), denoted by  $\dot{\alpha}_I$, is in the open interval $(0,1)$. Moreover, $\dot{\alpha}_I$ admits the following explicit formula:
\begin{align}\label{alpha_star_int}
    \dot{\alpha}_I=\frac{\sigma^2(v^I_l-v^I_u) }{\tau^2(b^I_u-b^I_l)+\sigma^2(v^I_l-v^I_u)},
\end{align}
and the minimum value of $r(\dot{\beta}^I_{\alpha})$ denoted by $r(\dot{\beta}^I_{\dot{\alpha}_I})$, admits the following explicit formula:
\begin{align*}
    r(\dot{\beta}^I_{\dot{\alpha}_I})= \frac{1}{2}\left[ \tau^2(\mathrm{tr}(\Sigma)-b_u)+\sigma^2 v^I_l -\frac{\sigma^4(v^I_l-v^I_u)^2 }{\tau^2(b^I_u-b^I_l)+\sigma^2(v^I_l-v^I_u)} \right] < r(\hat{\beta}^I).
\end{align*}
\end{theorem}
The proof of Theorem~\ref{Interpol1} is provided in~\ref{appA11}, and from it we conclude that $\dot{\beta}^I_{\dot{\alpha}_I}$ is a better estimator than both $\hat{\beta}^I$ and $\tilde{\beta}^I$. Furthermore, it is evident that $\dot{\alpha}_I$ increases with $\sigma^2$, thereby giving more significance to unlabeled data, similar to earlier conclusions in non-interpolating models. This behavior reinforces the intuition that unlabeled data become more valuable as the noise increases. The empirical findings detailed in Section \ref{Simulation-interpolators} corroborate the primary claim of Theorem~\ref{Interpol1}.

\section{Methodologies for choosing the mixing ratio}\label{Estimating-Methods}
The theoretical findings from the earlier sections demonstrate the advantages of incorporating unlabeled data through mixed-SSL with an appropriate mixing ratio. In practice, however, this ratio must be estimated from the available data. In this section, we introduce several methodologies to estimate appropriate mixing ratios in the various scenarios discussed in previous sections. We begin with the GLM case and describe estimators for $\dot{\alpha}$ and $\ddot{\alpha}$. We then turn to the interpolator model and propose a method for estimating $\dot{\alpha}_I$.

When considering the mixed estimator, $\dot{\beta}_{\alpha}$, we aim to estimate $\dot{\alpha}$, the minimizer of $\dot{r}_{\beta}(\alpha)$ (and approximately the minimizer of $R_{\beta}(\dot{\beta}_{\alpha})$). We do so via Equation (\ref{alp_dot_formula}), substituting estimates of $\sigma^2$, $v_l$, and $\dot{B}_{\beta}(\alpha=1)$. The terms $\dot{B}_{\beta}(\alpha=1)$ and $v_l$ can be determined using (\ref{dot_B_Def}) and (\ref{dot_V_Def}), by resampling $X$ from $Z$ and replacing $(\mu,D)$ with $(\breve{\mu},\breve{D})$, where $\breve{\mu}_i=g(x_i^t\breve{\beta})$ and $\breve{D}_{ii}=g'(x_i^t\breve{\beta})$. To estimate $\sigma^2$, we derive an approximate equality via a quadratic expansion of $RSS(\hat{\beta})=  \mid  \mid g(X\hat{\beta}) -Y  \mid  \mid _2^2$ around the true $\beta$:
\[ \mathbb{E}\left[RSS(\hat{\beta}) \mid X\right] \approx \sigma^2\left[\text{tr}\left( D  \right)-\text{tr}\left(X^tD^2X(X^tDX)^{-1}   \right) \right]. \]
The derivation of this approximation is explained in detail in Appendix~\ref{appC}. This results in the following approximate unbiased estimator of $\sigma^2$:
\[\hat{\sigma}^2= \frac{RSS(\hat{\beta})}{\text{tr}\left( \breve{D} \right)- \text{tr}\left(X^t\breve{D}^2X(X^t\breve{D}X)^{-1}   \right)}.\]
In the LM case, $\hat{\sigma}^2$ reduces to the standard unbiased estimator given by $\hat{\sigma}^2=RSS(\hat{\beta})/(n-p)$. 

We denote the data-driven estimate of $\dot{\alpha}$ by $\hat{\alpha}$, and the corresponding adaptive linear mixed estimator by $\dot{\beta}_{\hat{\alpha}}$. It is important to note that $\hat{\alpha}$ is a random variable that depends on the specific training sample. In Section \ref{Assimp-New} (Proposition \ref{ProbConv}), we focus on the LM under an asymptotic setting, and we establish that $\dot{\beta}_{\hat{\alpha}}$ performs comparably to the oracle estimator $\dot{\beta}_{\dot{\alpha}}$, which uses the true optimal mixing ratio as if it were known.

When considering the mixed estimator, $\ddot{\beta}_{\alpha}$, an explicit formula for $\ddot{\alpha}$ is unavailable. Consequently, we propose two practical methods for selecting the mixing ratio:
\begin{enumerate}
    \item Use $\hat{\alpha}$ from $\dot{\beta}_{\alpha}$ as a heuristic measure for $\ddot{\alpha}$. The estimator derived from this approach is denoted by $\Ddot{\beta}_{\hat{\alpha}}$.
    
    \item Perform a grid search over $\alpha \in (0,1)$ to identify the value $\tilde{\alpha}$ that minimizes the estimated $\Ddot{r}_{\beta}(\alpha)$, per the formulas (\ref{ddot_B_Def}) and (\ref{ddot_V_Def}), while using the unlabeled data and substituting $\beta$ with $\breve{\beta}$. The estimator obtained through this process is denoted by $\Ddot{\beta}_{\tilde{\alpha}}$.      
\end{enumerate}
In Section~\ref{GLM_Simulations}, we show through simulation that estimating $\dot{\alpha}$ and $\Ddot{\alpha}$ using the outlined method is feasible and leads to a significant enhancement in predictive performance.

When examining the linear mixed interpolator $\dot{\beta}^I_{\alpha}$, we aim to establish an estimator $\hat{\alpha}_I$ for the optimal mixing ratio $\dot{\alpha}_I$ as indicated in Equation (\ref{alpha_star_int}). We observe that the terms $b^I_l, b^I_u, v^I_l, v^I_u$ can be accurately calculated using the unlabeled data, whereas $\sigma^2$ and $\tau^2$ need to be estimated from the labeled data. We suggest a method for estimating $\sigma^2$ and $\tau^2$, and in Section~\ref{Simulation-interpolators}, we illustrate that it is feasible to apply this method to define the empirically-tuned mixed interpolator, $\dot{\beta}^I_{\hat{\alpha}_I}$, which outperforms the standard minimum-norm interpolator in all cases. Initially, we present an unbiased estimator for $\sigma^2$, assuming that $\tau^2$ is known.
\begin{prop}\label{P6} In the case of a true linear model with $p>n$, under Assumption \ref{A5}, the following provides an unbiased estimator of $\sigma^2$:
\begin{align}\label{F-Sigma}
   \hat{\sigma}^2 = \frac{Y^t(XX^t)^{-2}Y- \tau^2\text{tr}\left( (XX^t)^{-1} \right) }{\text{tr}\left( (XX^t)^{-2} \right)}.
\end{align}
\end{prop}

The proof of Proposition~\ref{P6} is provided in~\ref{appA12}, and we suggest using this unbiased estimator in order to define the empirically-tuned mixed interpolator $\dot{\beta}^I_{\hat{\alpha}_I(\tau)}$ in the known-$\tau$ scenario. In the general scenario where $\tau^2$ is unknown, it must be estimated together with $\sigma^2$. In this case, we suggest performing the following iterative calculation: 
\begin{align*}
    \hat{\tau}^2_{(0)}&=\frac{(\hat{\beta}^I)^t\Sigma \hat{\beta}^I}{\text{tr}(\Sigma)}   \hspace{1mm},\\
   \hat{\sigma}^2_{(t+1)} &=max\left\{ \frac{Y^t(XX^t)^{-2}Y- \hat{\tau}_{(t)}^2\text{tr}\left( (XX^t)^{-1} \right) }{\text{tr}\left( (XX^t)^{-2} \right)}  , 0 \right\},\\
    \hat{\tau}^2_{(t+1)} &= max\left\{ \left(\frac{\sum y_i^2}{n} - \hat{\sigma}^2_{(t+1)}\right)/\text{tr}\left(\Sigma \right)  , 0 \right\}.
\end{align*}
We note that typically only a few iterations are required to converge to stable estimates $(\hat{\sigma}^2, \hat{\tau}^2)$, which can be used to calculate the estimator $\hat{\alpha}_I$ of $\dot{\alpha}_I$. To the best of our knowledge, this estimation methodology is novel and can also be applied in other over-parameterized settings where the estimation of the noise and signal is required.

\section{Asymptotic properties of the mixed-SSL}\label{Assimp-New}
In this section, we examine the properties of the suggested mixed-SSL estimator for linear models, under an asymptotic setting where $p/n \to \gamma$ as $n,p\to \infty$, with fixed $\gamma>0$. In Section \ref{AsymOLS} we consider the case where $\gamma \in (0,1)$ and analyze the asymptotic behavior of $\eta=r(\dot{\beta}_{\dot{\alpha}}) / r(\hat{\beta})$, which is the relative mean reducible error between $\dot{\beta}_{\dot{\alpha}}$ and $\hat{\beta}$, in the random-$\beta$ scenario. In Section \ref{AsymInter} we consider the case where $\gamma \in (1,\infty)$ and analyze the asymptotic behavior of the relative mean reducible error in the over-parameterized model, $\eta^I=r(\dot{\beta}^I_{\dot{\alpha}_I})/ r(\hat{\beta}^I)$. We aim to show that both $\eta$ and $\eta^I$ remain strictly smaller than $1$ as $n$ and $p$ diverge, confirming that the benefit of unlabeled data persists even as the labeled sample grows.


\subsection{Asymptotic properties in non-interpolating LM}\label{AsymOLS}
We now adopt the setting of a random coefficient vector for the non-interpolating LM, for brevity and simplicity. We assume that $\beta$ is drawn independently, such that $\mathbb{E}\left[ \beta \right] = \textbf{0}_p$ and $\mathbb{E}\left[ \beta \beta^t \right] = \tau^2 I_p$. In this scenario, the mean reducible error of $\dot{\beta}_{\alpha}$, averaged over $P_\beta$, is:
\begin{align*}
r(\dot{\beta}_{\alpha})& = \mathbb{E}_{\beta}\left[r_{\beta}(\dot{\beta}_{\alpha})\right] =  \mathbb{E}_{\beta}\left[B_{\beta}(\dot{\beta}_{\alpha})\right]+\mathbb{E}_{\beta}\left[V_{\beta}(\dot{\beta}_{\alpha})\right] \\ &= \frac{1}{2}\tau^2 b_u +\frac{\sigma^2}{2} \left(\alpha^2 v_u +(1-\alpha)^2v_l +2\alpha(1-\alpha) v_u  \right),
\end{align*}
where:
\[b_u= \frac{1}{n}\alpha^2 \text{tr}\left( \mathbb{E}_X \left[ \left(X^tX- n\overline{X} \hspace{1mm} \overline{X}^t  -H\right)  \left( H^{-1} (X^tX- n\overline{X} \hspace{1mm} \overline{X}^t ) -I  \right)^t \right]  \right),\]
and $v_l$, as defined in (\ref{dot_V_Def}), can now be expressed as $\frac{1}{n}\text{tr}\left( \mathbb{E}_X\left[ (X^tX)^{-1} \right] \mathbb{E}_X\left[X^tX \right] \right)$ due to the condition $D=I_n$.

We note that in this case, the bias term simplifies to $\frac{1}{2}\tau^2 b_u$, where $b_u$ depends only on $P_x$ and can be estimated precisely from the unlabeled data. Using this identity, we adapt Theorem \ref{P2n} as follows:
\begin{align}\label{E_alpha_dot}
    \dot{\alpha}=\frac{\sigma^2(v_l-v_u) }{\tau^2b_u+\sigma^2(v_l-v_u)} ,
    \end{align}
    \begin{align}\label{E_R_dot}
    r(\dot{\beta}_{\dot{\alpha}})= \frac{1}{2} \left[ \sigma^2v_l -\frac{\sigma^4(v_l-v_u)^2 }{\tau^2b_u+\sigma^2(v_l-v_u)}\right],
\end{align}
and as a consequence, we can derive the following representation for $\eta$:
\begin{align}\label{Eeta}
    \eta =  \frac{r(\dot{\beta}_{\dot{\alpha}})}{r(\hat{\beta})} = 1- \frac{\sigma^2(v_l-v_u)^2}{ \tau^2 v_lb_u+\sigma^2v_l(v_l-v_u)}.
\end{align}
This expression shows how $\eta$ depends jointly on $\sigma^2$, $\tau^2$, and $P_x$. To thoroughly examine the asymptotic properties of $\eta$, we investigate an asymptotic scenario where $p/n \to \gamma\in(0,1)$ as $n,p\to \infty$, and the covariates' distribution adheres to the following assumption.
\begin{Assump}\label{A01} 
The distribution $P_x$ satisfies $\mathbb{V}ar(x)=\Sigma$, where $\Sigma$ is (a sequence of) invertible $p\times p$ matrices, and $\text{tr}(\Sigma)\to c^2$ for some $c\in \mathbb{R}$, as $p\to \infty$.
\end{Assump}

The following proposition describes the asymptotic behavior of $\eta$ and $\dot{\alpha}$ under the conditions of Assumption~\ref{A01}, and also $x\sim MN(\textbf{0}_p,\Sigma)$, which represents a multivariate normal distribution with mean $\textbf{0}_p$ and covariance matrix $\Sigma$. 

\begin{prop} \label{GassianAssym1}
    Consider the asymptotic setting $p/n \to \gamma\in(0,1)$ as $n,p\to \infty$, and assume that $x\sim MN(\textbf{0}_p,\Sigma)$, where $\Sigma$ follows the conditions of  Assumption~\ref{A01}, then in the limit as $n,p \to \infty$ we obtain:
    \begin{align*}
    \eta \to  \eta_{\infty} = 1- \frac{\gamma^4\sigma^2}{(1-\gamma)\gamma^2\tau^2c^2+\gamma^3 \sigma^2}  <1  \hspace{2mm};\hspace{2mm}
    \dot{\alpha}\to \dot{\alpha}_{\infty}=\frac{\gamma^2\sigma^2}{(1-\gamma)\gamma \tau^2c^2+\gamma^2 \sigma^2}.
\end{align*}
\end{prop}
Appendix \ref{appA5} contains the proof of Proposition~\ref{GassianAssym1}, indicating that the limit of $\eta$ is strictly smaller than $1$ for any $\gamma\in(0,1)$. This confirms that unlabeled data retain value asymptotically. Additionally, insights can be gained regarding the impact of dimensionality on the relative advantage of using unlabeled data. As $\gamma$ approaches $0$, $\eta$ tends toward $1$ and $\dot{\alpha}$ toward $0$, indicating that it is not advantageous to use unlabeled data. In contrast, as $\gamma$ approaches $1$, $\eta$ tends toward $0$ and $\dot{\alpha}$ toward $1$, suggesting that the use of unlabeled data is highly beneficial. The key insight from Proposition \ref{GassianAssym1} about the asymptotic behavior of $\eta$ can be generalized to encompass a broader range of covariate distributions. The following theorem describes the asymptotic behavior of $\eta$ and $\dot{\alpha}$ under the general covariates' generating mechanism of Assumption~\ref{AMechanisim_n}.

\begin{theorem}\label{GenAssym}
        Consider the asymptotic setting $p/n \to \gamma\in(0,1)$ as $n,p\to \infty$, and assume that Assumptions~\ref{A0n}, \ref{AMechanisim_n}, and \ref{A01} hold. Then for any $\epsilon > 0$, there exists $n(\epsilon)$ such that: 
    \begin{align*}
    \eta  \leq \eta_{\infty}+\epsilon \hspace{2mm} \text{and} \hspace{2mm}
    \dot{\alpha} \geq \dot{\alpha}_{\infty} -\epsilon \hspace{2mm},\hspace{2mm} \forall n>n(\epsilon).
\end{align*}
\end{theorem}

The proof of Theorem~\ref{GenAssym} can be found in Appendix \ref{appA6}, and from it we conclude that under broad distributional conditions, the limiting $\eta$ lies in $(0,\eta_{\infty}]$, showing that $\dot{\beta}_{\dot{\alpha}}$ significantly outperforms $\hat{\beta}$. Consider the scenario where $\tau^2$ is known, but we need to estimate $\sigma^2$ in order to evaluate the best mixing ratio $\dot{\alpha}$. Let $\dot{\beta}_{\hat{\alpha}(\tau)}$ denote the estimator that uses known $\tau^2$ together with $\hat{\sigma}^2$ of $\sigma^2$ to estimate $\dot{\alpha}$ via Equation (\ref{E_alpha_dot}). The following proposition indicates that we should expect a comparable predictive performance between $\dot{\beta}_{\hat{\alpha}(\tau)}$ and $\dot{\beta}_{\dot{\alpha}}$.

\begin{prop}\label{ProbConv}
    Consider the asymptotic setting described above and assume that the distribution $P_x$ satisfies $b_u\to \gamma c^2$, and $v_l\to \gamma/(1-\gamma)$ as $n\to \infty$, then the random difference between the predictions made using $\dot{\beta}_{\hat{\alpha}(\tau)}$ and $\dot{\beta}_{\dot{\alpha}}$ satisfies:
    \[x_0^t\dot{\beta}_{\dot{\alpha}}  -x_0^t\dot{\beta}_{\hat{\alpha}(\tau)} \overset{P}{\rightarrow} 0 \hspace{2mm}, \text{as } n\to \infty.\]
\end{prop}
The proof of Proposition~\ref{ProbConv} can be found in Appendix \ref{appA7}. Proposition \ref{ProbConv} offers an understanding of the similarity between the data-driven estimator $\dot{\beta}_{\hat{\alpha}(\tau)}$ and the oracle estimator $\dot{\beta}_{\dot{\alpha}}$ when $\tau^2$ is known. On the other hand, if $\tau^2$ is unknown, the following estimators can be used, as proposed in \cite{yuval2022semi}:
\[ \hat{\tau}^2= max\left\{ \left(\frac{\sum_{i=1}^n y_i^2}{n} - \hat{\sigma}^2\right)/\text{tr}\left(\mathbb{E}\left[xx^t \right] \right)  , 0 \right\}  ,\hspace{1mm} \text{where} \hspace{1mm}\hat{\sigma}^2=\frac{RSS(\hat{\beta})}{n-p}.\]
In this case, the variance of $\hat{\tau}^2$ is $2\tau^4 \text{tr}(\Sigma^2)/(\text{tr}(\Sigma))^2$, which does not vanish with $n$ in general (depending on the structure of $\Sigma$) and therefore the estimator $\hat{\alpha}$ is not consistent. Consequently, $\dot{\beta}_{\hat{\alpha}}$ may deviate from $\dot{\beta}_{\dot{\alpha}}$ and yield a different predictive error. Interestingly, in practice, estimating $\tau^2$ often outperforms using the true value (Section~\ref{OLS_Simulations}, Figure~\ref{OLS3}). This phenomenon can be explained by considering that in $\dot{\beta}_{\dot{\alpha}}$, we adjust the mixing ratio to the specific data realization at hand, which corresponds to the specific instance of $\beta$.

\subsection{Asymptotic properties in mixed linear interpolating models}\label{AsymInter}
The relative reducible predictive error between the oracle-tuned interpolator, $\dot{\beta}^I_{\dot{\alpha}_I}$, and the supervised interpolator, $\hat{\beta}^I$, is:
\begin{align*}
    \eta^I=  \frac{r(\dot{\beta}^I_{\dot{\alpha}_I})}{r(\hat{\beta}^I)} = 1- \frac{\sigma^4(v^I_l-v^I_u)^2}{   \left(\tau^2(b^I_u-b^I_l)+\sigma^2(v^I_l-v^I_u) \right)\left( \tau^2 b^I_l +\sigma^2v^I_l \right)}.
\end{align*}
Detailed analysis of the characteristics of the quantities $(v^I_l,v^I_u,b^I_u,b^I_l)$ and the estimator $\hat{\alpha}$, in different settings, especially with respect to $P_x$, is beyond the scope of this paper. For the sake of brevity and to highlight the advantage of the semi-supervised interpolator over the supervised one, we consider the following distributional assumption.
\begin{Assump}\label{AHighDim}
    The distribution $P_x$ satisfies $x\sim MN(\textbf{0},\Sigma)$, where $\Sigma$ is (a sequence of) $p\times p$ diagonal matrices with entries $\Sigma_{jj}=c^2_1(n)$ for $1\leq j \leq \tilde{p}$ and $\Sigma_{jj}=c^2_2(n)$ for $\tilde{p} < j \leq p$.  Moreover, the asymptotic setup satisfies the following: 
    \begin{align*}
        p/n \to \gamma\in(1,\infty) \hspace{2mm}&;\hspace{2mm} \tilde{p}/n \to \tilde{\gamma} \in  (1,\gamma)\\
        \tilde{p} c^2_1(n) = \sum_{j=1}^{\tilde{p}} \Sigma_{jj} \to c^2  \hspace{2mm}&;\hspace{2mm} (p-\tilde{p}) c^2_2(n) = \sum_{j=\tilde{p}}^{p} \Sigma_{jj} \to 0,
    \end{align*}
    as $n,p,\tilde{p} \to \infty$, for some constant $c\in \mathbb{R}$.
\end{Assump}
The setting of Assumption~\ref{AHighDim} highlights the benefit of using the available unlabeled data, which leads to the inference of the unbalanced structure of $\Sigma$ and the subsequent adjustment of the linear interpolator. Moreover, in this setting, we can find explicit expressions for the limiting values of $v^I_l$, $v^I_u$, $b^I_l$ and $b^I_u$, and also for $\dot{\alpha}_I$ and $\eta^I$. The following theorem outlines the asymptotic properties of the mixed-SSL interpolator, indicating that $\eta^I$ remains strictly smaller than $1$ as the sample size grows to infinity.

\begin{theorem}\label{AsymptoticInterpolators}
        Under the conditions of  Assumption~\ref{AHighDim}, as $n\to \infty$, the following limiting values exist:
    \[v^I_l  \to \frac{1}{\tilde{\gamma}-1} \hspace{2mm};\hspace{2mm} b^I_l  \to c^2\frac{1}{\tilde{\gamma}} \hspace{3mm};\hspace{3mm} v^I_u \to \frac{1}{\gamma-1} \hspace{2mm};\hspace{2mm} b^I_u  \to c^2\frac{1}{\gamma},\]
    \begin{align*}
    \eta^I &\to  \eta^I_{\infty} = 1- \frac{(\gamma-\Tilde{\gamma})\sigma^4}{(\gamma-1)^2(\Tilde{\gamma}-1)^2 \left[  \frac{\tau^2c^2}{\gamma \Tilde{\gamma}} +\frac{\sigma^2}{(\gamma-1)(\Tilde{\gamma}-1)} \right]  \left[  \frac{\tau^2c^2(\Tilde{\gamma}-1)}{\Tilde{\gamma}} +\frac{\sigma^2}{\Tilde{\gamma}-1} \right]},\\
    \dot{\alpha}_I &\to \dot{\alpha}_{I,\infty} =\frac{\sigma^2}{(\gamma-1)(\Tilde{\gamma}-1) \left[  \frac{\tau^2c^2}{\gamma \Tilde{\gamma}} +\frac{\sigma^2}{(\gamma-1)(\Tilde{\gamma}-1)} \right]}.
\end{align*}
\end{theorem}

The proof of Theorem~\ref{AsymptoticInterpolators} is provided in Section~\ref{appA13}, and from it, we conclude that the predictive error of $\dot{\beta}^I_{\dot{\alpha}_I}$ is strictly lower than the one of $\hat{\beta}^I$ as long as $\tilde{\gamma}<\gamma$. Moreover, we can see that $\eta^I$ increases with $\tilde{\gamma}$ and is equal to $1$ when $\tilde{\gamma}=\gamma$. On the other hand, $\eta^I$ decreases with $\gamma$ and approaches $0$ when $\tilde{\gamma}$ approaches $1$. While we are not able to exactly identify the characteristics of the estimates $(\hat{\sigma}^2, \hat{\tau}^2)$, the typical behavior we observed in our simulations is that if $\tau^2$ is known, the variance of $\hat{\sigma}^2$ is of the order $1/n$. Moreover, if $\sigma^2$ is known, the variance of $\hat{\tau}^2$ vanishes as n diverges, since $\text{tr}(\Sigma^2)$ approaches zero. Therefore, we expect that the empirically tuned estimators $\dot{\beta}^I_{\hat{\alpha}_I}$ and $\dot{\beta}^I_{\hat{\alpha}_I(\tau)}$ will perform similarly to $\dot{\beta}^I_{\dot{\alpha}_I}$. In Section \ref{Simulation-interpolators}, we demonstrate that this is indeed a typical behavior.

\section{Results of synthetic data simulations}\label{Results} 
In this section, we provide an empirical analysis of all the estimators discussed in Sections \ref{semi-supervised-GLM} and \ref{Interpolators-Sec-New}, using simulated data. This analysis substantiates the theoretical findings from the earlier sections.

\subsection{Simulations for LM} \label{OLS_Simulations}
We present an empirical analysis of the predictive performance of the standard LM along with all SSL variants discussed in Section~\ref{semi-supervised-GLM}, under two different data-generating mechanisms, as follows:
\begin{enumerate}
\item \textit{Constant}-$\beta$. $\mathbb{E}[y \mid x] = x^t\beta \hspace{2mm},\hspace{2mm} \beta=1.5\cdot\mathbf{1}_p$, where $n=100$ and $p=50$, across various values of $\sigma^2$. 
\item \textit{Random}-$\beta$. $\mathbb{E}[y \mid x,\beta] = x^t \beta$, where $\beta \sim MN(\textbf{0},I_p\tau^2) $ is generated with each training set, across various values of $n$, with $p=0.5n$ and constants $\sigma^2=25$ and $\tau^2=1$ used consistently.   
\end{enumerate}
In every scenario, considering any configuration of $(n,p,\sigma^2)$, the covariates are drawn from a multivariate normal distribution, $ MN(\textbf{0}_p, \Sigma)$ where $\Sigma$ is $p \times p$ block-diagonal, containing five blocks such that all variables in each block have pairwise correlation $\rho=0.9$, such that $\text{tr}(\Sigma)=25$. Initially, we conduct an evaluation of the relevant statistics that depend only on the distribution of the covariates, such as $v_l$, $v_u$, $H$, based on a large fixed dataset $Z$ of $m=5\times 10^4$ unlabeled observations $z \sim MN(\textbf{0}_p, \Sigma)$. We also use a large fixed dataset $\mathcal{Z}$ of $N=5\times 10^4$ unlabeled observations to evaluate the oracle mixing ratios, $\dot{\alpha}$ and $\Ddot{\alpha}$, based on the knowledge of the true parameters of the models. We then generate $K=5,000$ random training sets of labeled observations ($X,Y$) where the rows of $X$ are randomly sampled from $MN(\textbf{0}_p, \Sigma)$ and $Y=X\beta+\epsilon_n$, where $\epsilon_n \sim MN(\textbf{0}_n, \sigma^2I_n)$. For each of the training sets, we fit the supervised and semi-supervised LM estimators $(\hat{\beta},\Breve{\beta},\dot{\beta}_{\hat{\alpha}})$, as well as the oracle estimator $\dot{\beta}_{\dot{\alpha}}$.

\begin{figure}[ht!] 
\centering
\includegraphics[width=9cm]{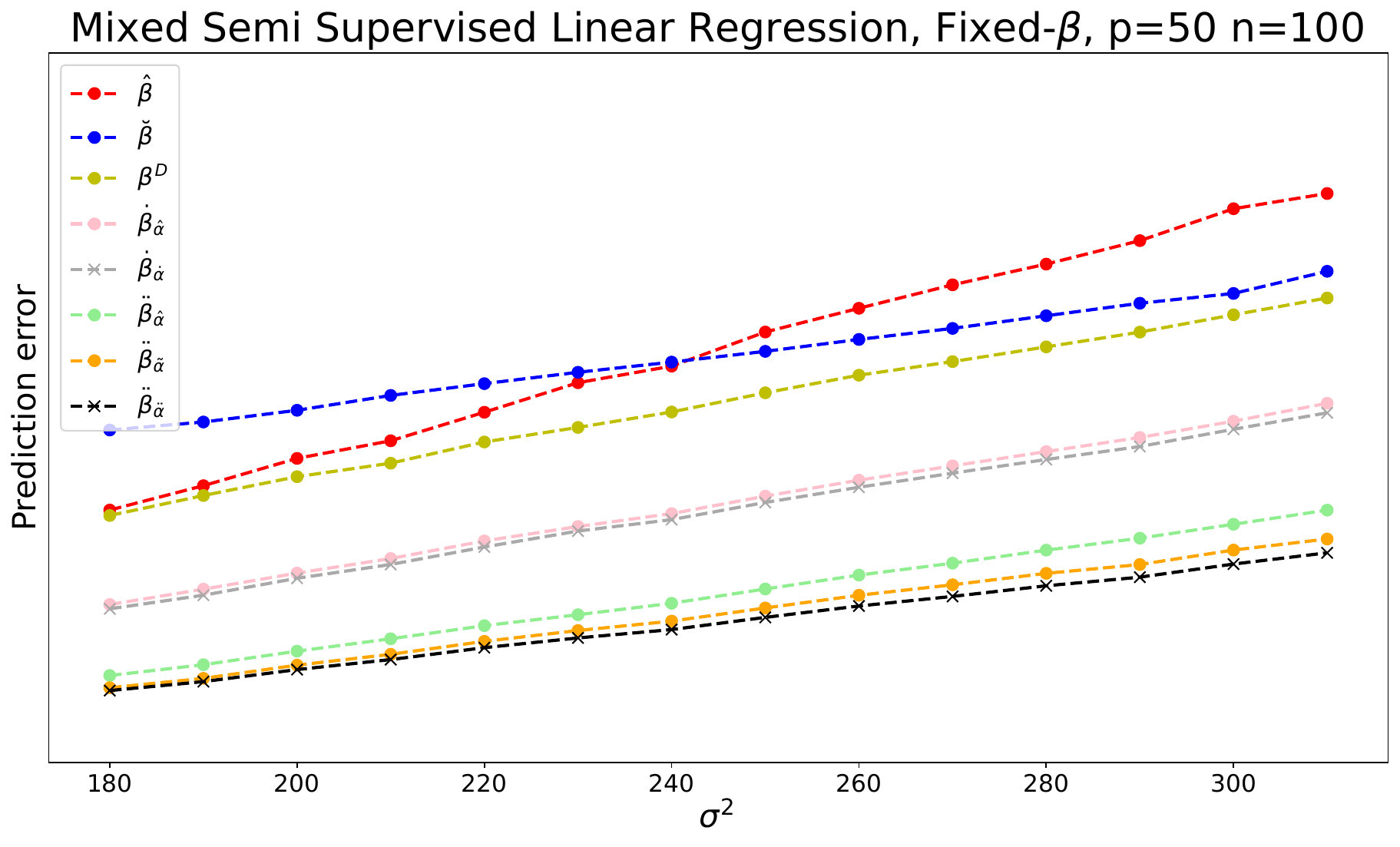}
\caption{Mean reducible predictive errors of LM with constant-$\beta$ mechanism, for various values of $\sigma^2$, with $n=100$ and $p=50$ (see text for details).
}
\label{OLS1}
\end{figure}

\begin{figure}[ht!] 
\centering
\includegraphics[width=9cm]{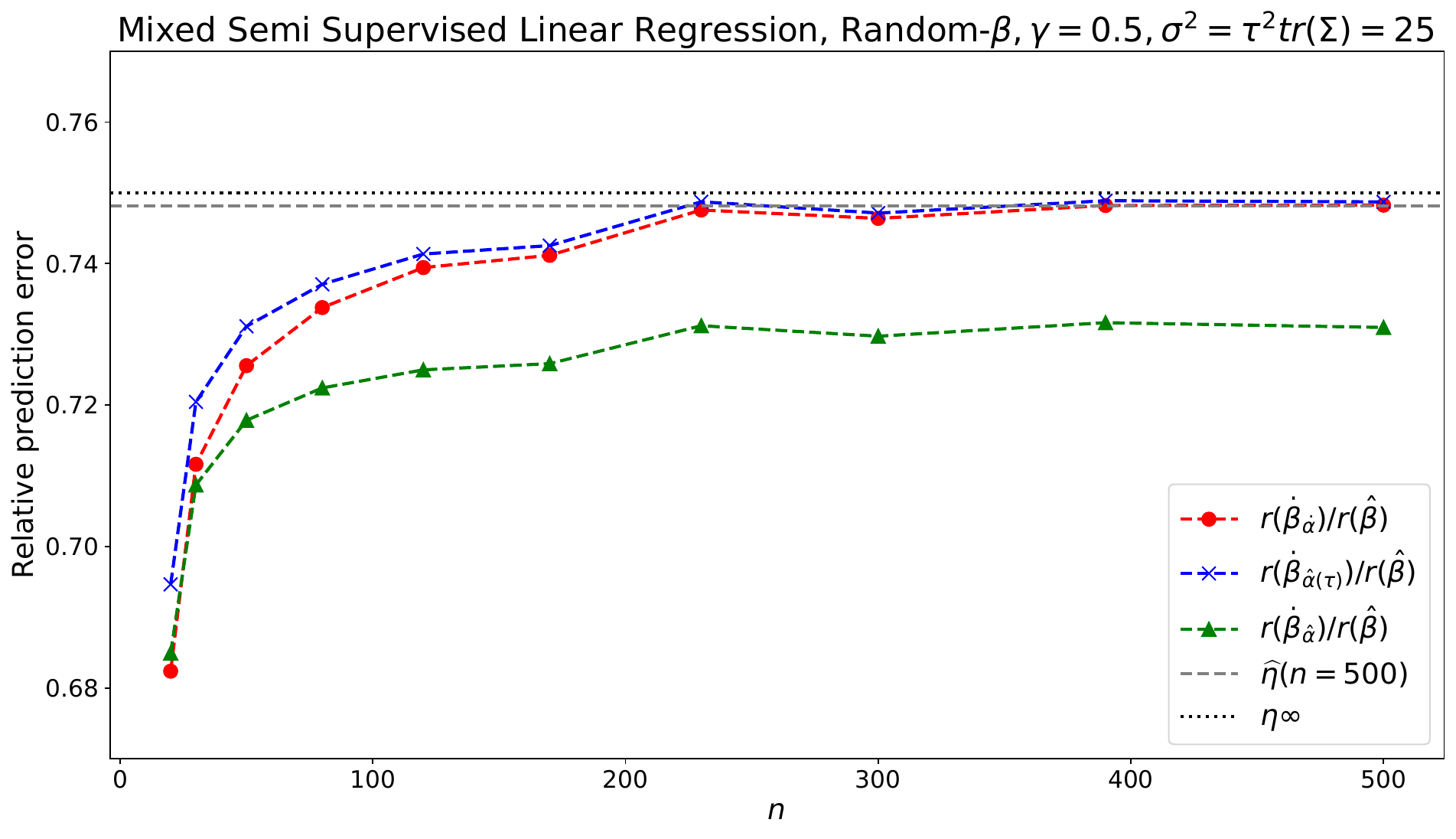}
\caption{Relative reducible predictive errors of the random-$\beta$ mechanism for various values of $n$, with $\gamma=0.5$, and $\text{tr}(\Sigma)=\sigma^2=25$ (see text for details).
}
\label{OLS3}
\end{figure}

Specifically, in the case of the \textit{Constant}-$\beta$ generating mechanism, we also fit the semi-supervised estimators $(\beta^D,\Ddot{\beta}_{\hat{\alpha}},\Ddot{\beta}_{\tilde{\alpha}})$, as well as the oracle estimator $\ddot{\beta}_{\ddot{\alpha}}$. The mean predictive error across the dataset $\mathcal{Z}$ is calculated for each fitted estimator (as per Equation (\ref{Rdef1}), given the training sample $T=(X,Y)$) and then averaged over the $K$ training samples. The outcome is eight curves that describe the predictive error, changing with $\sigma^2$ (Figure \ref{OLS1}). In the \textit{Random}-$\beta$ generating mechanism, the estimator $\dot{\beta}_{\hat{\alpha}(\tau)}$ is also fitted, and the mean predictive error is computed conditioned on the sampled $\beta$ for every training sample. The outcome is three curves that describe the reducible errors associated with $\dot{\beta}_{\dot{\alpha}}$, $\dot{\beta}_{\hat{\alpha}}$, and $\dot{\beta}_{\hat{\alpha}(\tau)}$, relative to that of $\hat{\beta}$, changing with $n$ (Figure \ref{OLS3}). Formally, for each specified value of $n$, we calculated the following empirical ratios: $r(\dot{\beta}_{\dot{\alpha}})/r(\hat{\beta})$, $r(\dot{\beta}_{\hat{\alpha}(\tau)})/r(\hat{\beta})$, and $r(\dot{\beta}_{\hat{\alpha}})/r(\hat{\beta})$, arranged in a decreasing order according to the knowledge of the true parameters.

In Figure \ref{OLS1}, we can see that the mixed-SSL is very useful in reducing the mean predictive error. In particular, the estimator $\Ddot{\beta}_{\tilde{\alpha}}$ outperforms all other estimators, excluding the oracle estimator $\ddot{\beta}_{\ddot{\alpha}}$, with a benefit that increases with $\sigma^2$. We can see that the estimator $\ddot{\beta}_{\hat{\alpha}}$ is also very useful despite the fact that it uses the mixing ratio aimed at $\dot{\beta}_{\hat{\alpha}}$. The oracle estimators are only slightly better than $\dot{\beta}_{\hat{\alpha}}$ and $\Ddot{\beta}_{\tilde{\alpha}}$ respectively, implying that there is little advantage in knowing the exact optimal mixing ratio compared to estimating it. All variances of pairwise mean differences are tiny; hence, any difference larger than the marker's size is significant at the $0.05$ level in a paired $t$-test. 

In Figure \ref{OLS3} we mark the theoretical limit of $\eta$ in this setting, $\eta_{\infty}$, by a dotted horizontal black line, and the estimated value of $\eta$ based on the dataset $\mathcal{Z}$ with $n=500$, by a dashed horizontal gray line. In the results, we can see that the relative reducible predictive error of both $\dot{\beta}_{\dot{\alpha}}$ and $\dot{\beta}_{\hat{\alpha}(\tau)}$ converges to the expected limiting value of $1-0.5^2$, which supports the theoretical analysis. Interestingly, the predictive performance of the most realistic estimator, $\dot{\beta}_{\hat{\alpha}}$, converges to a significantly lower value, supporting the conjecture that the non-vanishing variance of the estimator $\hat{\tau}^2$ leads to a better estimation of the mixing ratio, compared to knowing the real value of $\tau^2$.




\subsection{Simulations for GLM}\label{GLM_Simulations}
We perform an experiment similar to that for the LM case, with the \textit{Constant}-$\beta$ mechanism, but now using the ELU function \cite{clevert2015fast} as the link function $g$: $\mathbb{E}[y \mid x] = \min\left\{e^{x^t\beta}-1,max\left(0,x^t\beta\right)  \right\}$, where $ \beta=2\cdot \mathbf{1}_p$ and $p=10$. The covariates generating mechanism is the same as in the LM scenario, and a large fixed dataset $\mathcal{Z}$ consisting of $N=5\times 10^4$ unlabeled observations is utilized for determining the oracle mixing ratios and assessing predictive performance. However, for this case, we generate $K=5000$ different random training samples $(X,Y,Z)$, each containing $n=50$ labeled observations ($X$ and $Y$), and $m=5000$ unlabeled observations ($Z$), spanning different values of $\sigma^2$. This selection of $m$ is made to ensure that computational resources remain manageable in this context. Furthermore, this configuration highlights that a relatively small quantity of unlabeled data can also be practically utilized.

For each of the training sets, we fit the estimators $\hat{\beta}$ and $\Breve{\beta}$. Then, by sub-sampling $Z$, at the fitted vector $\Breve{\beta}$, we evaluate $\hat{\sigma}^2$ and all other unsupervised statistics in order to obtain estimates of $\dot{\alpha}$ and $\Ddot{\alpha}$, denoted by $\hat{\alpha}$ and $\tilde{\alpha}$ respectively, and use them to fit the mixed estimators $(\dot{\beta}_{\hat{\alpha}},\ddot{\beta}_{\hat{\alpha}},\ddot{\beta}_{\tilde{\alpha}})$. The method for fitting $\ddot{\beta}_{\hat{\alpha}}$ and $\ddot{\beta}_{\tilde{\alpha}}$ is summarized by the following Newton-Raphson update step, where $\alpha \in \{\hat{\alpha},\tilde{\alpha} \}$ respectively:
\[\beta^{(t+1)} =  \beta^{(t)} -\left(S_{\alpha}^{(t)}    \right)^{-1} \left(\alpha\zeta_{m}^{(t)} + (1-\alpha)\zeta_{n}^{(t)}  \right),  \]
where,
\begin{align*}
    S_{\alpha}^{(t)}&= \alpha H_{m}^{(t)} + (1-\alpha)H_{n}^{(t)} \hspace{2mm};\hspace{2mm}
     H_{m}^{(t)}= \frac{1}{m}Z^t D_{m}^{(t)} Z \hspace{2mm};\hspace{2mm} \left[D_{m}^{(t)}\right]_{ii}= g'(z_i^t\beta^{(t)}), \\
     H_{n}^{(t)} &= \frac{1}{m}X^t D_{n}^{(t)} X \hspace{2mm};\hspace{2mm} \left[D_{n}^{(t)}\right]_{ii}= g'(x_i^t\beta^{(t)}),\\
     \zeta_{m}^{(t)}&= \frac{1}{m} Z^tg(Z\beta^{(t)})  -\overline{ Z}\cdot \overline{Y} -  \widehat{\mathbb{C}ov}(X,Y) \hspace{2mm};\hspace{2mm}  \zeta_{n}^{(t)}= \frac{1}{n} X^t \left(g(X\beta^{(t)})-Y\right).     
\end{align*}
We also fit the oracle estimators $\dot{\beta}_{\dot{\alpha}}$ and $\ddot{\beta}_{\ddot{\alpha}}$, based on the oracle mixing ratios for $\dot{r}(\alpha)$ and $\Ddot{r}_{\beta}(\alpha)$ respectively, from the first step of the simulation. The mean predictive error over the dataset $\mathcal{Z}$ is evaluated for every fitted estimator (according to Equation (\ref{Rdef1})), and averaged over the $K$ training samples. The outcome is eight curves that describe the predictive error $R$, changing with $\sigma^2$ for each of the estimators.

\begin{figure}
\centering
\includegraphics[width=9cm]{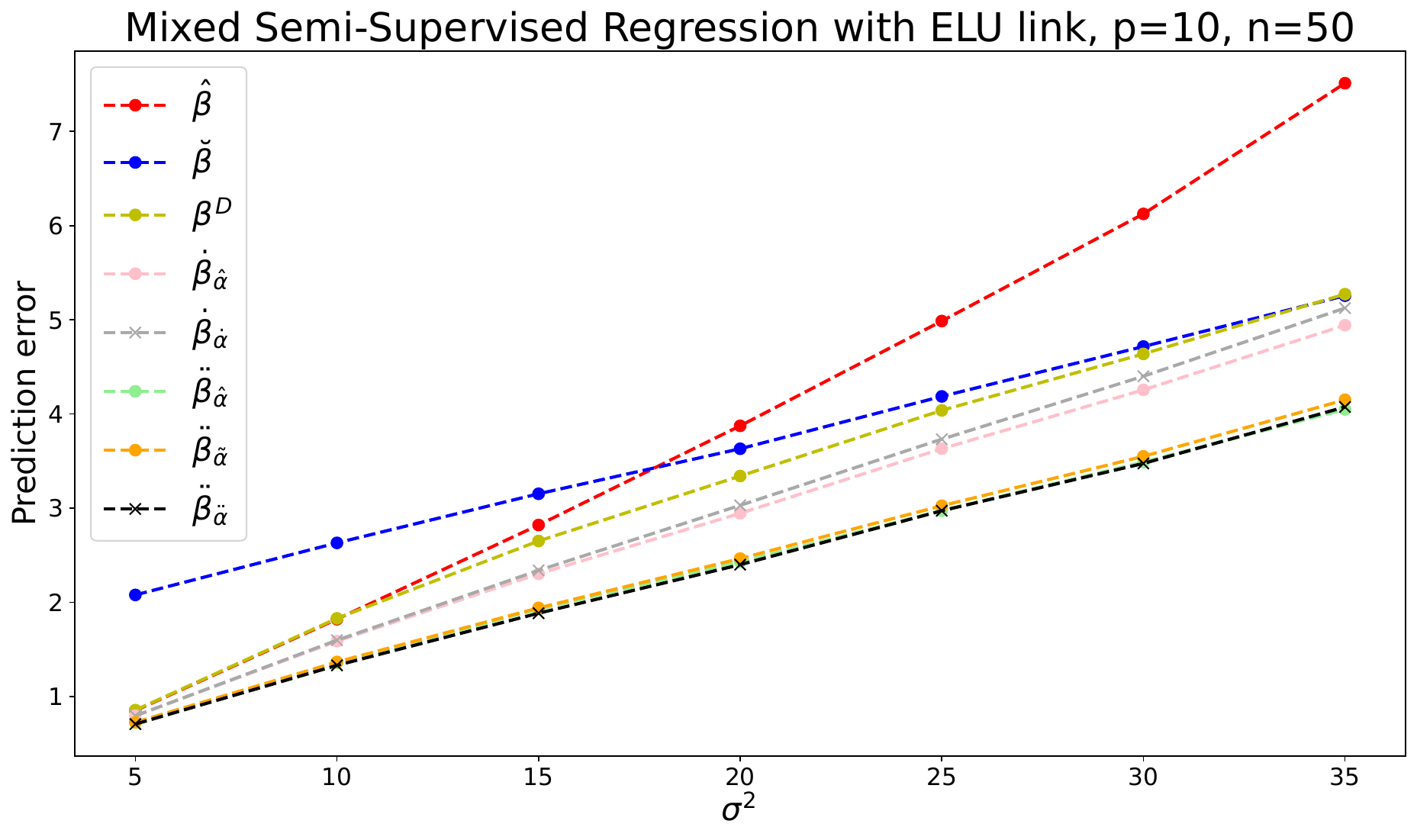}
\caption{Mean predictive errors of ELU model, for various values of $\sigma^2$, with $n=50$ and $p=10$ (see text for details).}
\label{GLM1}
\end{figure}

In the results (Figure \ref{GLM1}), we can see that mixed-SSL is useful in reducing the predictive error, and that the data-driven estimators $\ddot{\beta}_{\tilde{\alpha}}$, $\ddot{\beta}_{\hat{\alpha}}$ and $\dot{\beta}_{\hat{\alpha}}$ work well in practice. In particular, it is evident that $\dot{\beta}_{\hat{\alpha}}$ slightly outperforms the oracle estimator $\dot{\beta}_{\dot{\alpha}}$. Additionally, $\ddot{\beta}_{\hat{\alpha}}$ shows marginally better performance compared to $\ddot{\beta}_{\tilde{\alpha}}$ and is comparable to the oracle estimator $\ddot{\beta}_{\ddot{\alpha}}$, even though it utilizes the estimate of $\dot{\alpha}$ as the mixing ratio. Any difference greater than the marker size is significant at the $0.05$ level in a paired $t$-test.

\begin{figure}[ht] 
\centering
\includegraphics[width=9cm]{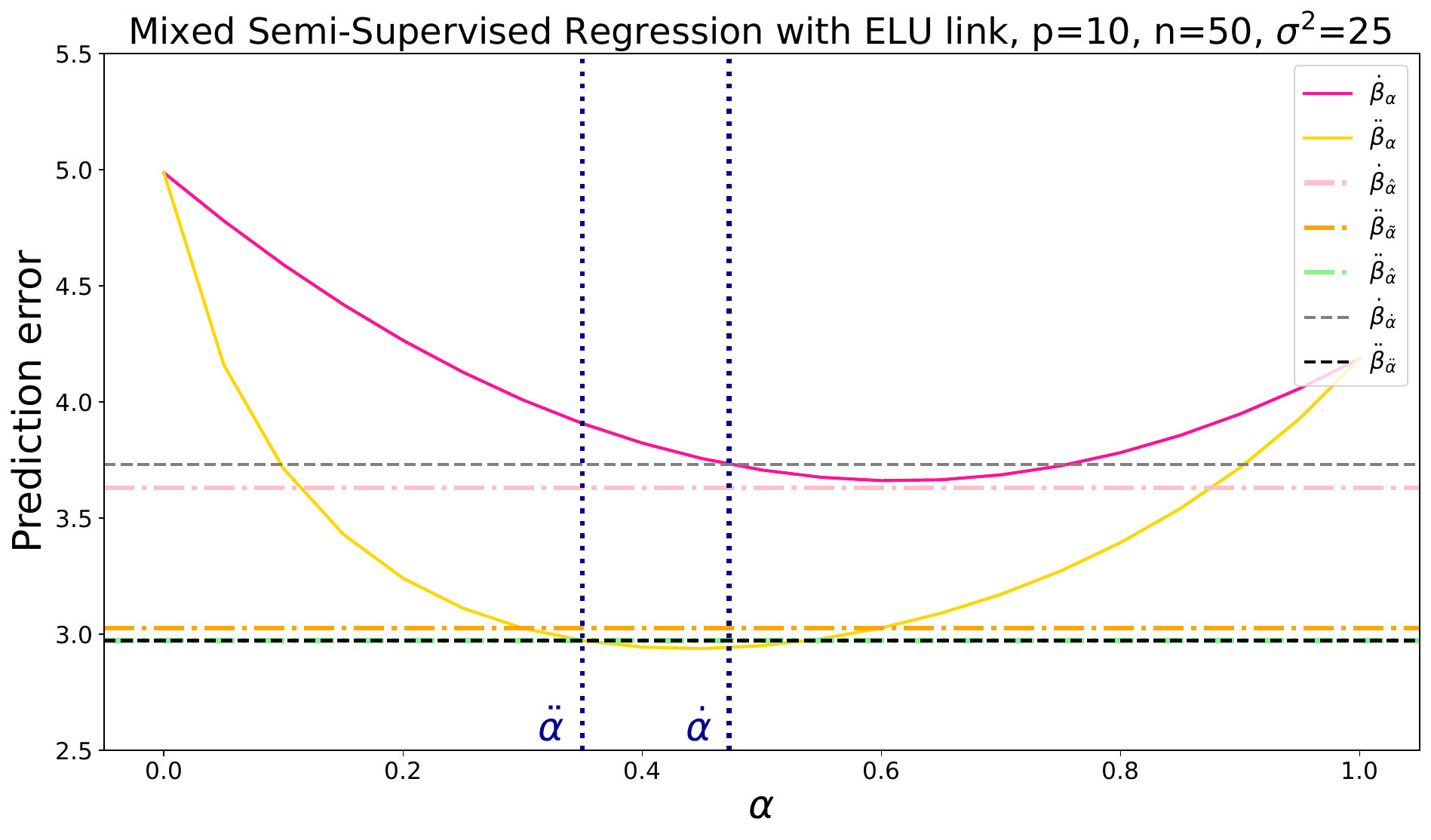}
\caption{Mean predictive errors of ELU model, for various values of $\alpha$, with $\sigma^2=25$, $n=50$ and $p=10$ (see text for details).}
\label{GLM2}
\end{figure}

In Figure \ref{GLM2} we evaluate the mean predictive error of $\dot{\beta}_{\alpha}$ and $\Ddot{\beta}_{\alpha}$ for various values of $\alpha$ in the interval $[0,1]$, specifically for $\sigma^2=25$. The outcome is two curves that describe the predictive errors, $R_{\beta}(\dot{\beta}_{\alpha})$ and $R_{\beta}(\Ddot{\beta}_{\alpha})$, changing with $\alpha$. For reference, we mark the oracle evaluations of $\dot{\alpha}$ and $\Ddot{\alpha}$ in vertical lines, and the predictive error of the other mixed estimators in horizontal lines. The results demonstrate that $R_{\beta}(\Ddot{\beta}_{\alpha})<R_{\beta}(\dot{\beta}_{\alpha})$ for all values of $\alpha$, corroborating the main observation of Section \ref{semi-supervised-GLM} regarding the relationship between the two estimators. It is notable that $R_{\beta}(\dot{\beta}_{\hat{\alpha}})$ is less than $R_{\beta}(\dot{\beta}_{\alpha})$ for any fixed $\alpha$, implying that $\hat{\alpha}$, as a random variable dependent on the training sample, results in improved tuning of the mixing ratio. More importantly, we can see that both $\dot{\alpha}$ and $\Ddot{\alpha}$ (especially) are very good approximations for determining the best mixing ratios of $\dot{\beta}_{\alpha}$ and $\Ddot{\beta}_{\alpha}$ respectively, despite the quadratic approximation and the fact that, by definition, they are the minimizers of the $\dot{r}_{\beta}(\alpha)$ and $\ddot{r}_{\beta}(\alpha)$ respectively, and not of $R_{\beta}(\dot{\beta}_{\alpha})$ and $R_{\beta}(\ddot{\beta}_{\alpha})$. The differences between $R_{\beta}(\ddot{\beta}_{\alpha})$ and $R_{\beta}(\dot{\beta}_{\alpha})$ are significant for $\alpha \in [0.05,0.95]$.


\subsection{Simulation for linear interpolators}\label{Simulation-interpolators}
We perform an experiment similar to that in the case of non-interpolating LM, with random coefficients, $\beta \sim MN(\textbf{0},I_p\tau^2) $, where $\tau^2=1$, $p=100$, $n=50$, and $\Sigma$ is a diagonal matrix with components $\Sigma_{ii}=1$ for $i\leq 0.8p$, and $\Sigma_{ii}=1/n$ for $i>0.8p$. For each of the training sets, we fit the interpolators $(\hat{\beta}^I,\tilde{\beta}^I)$, and then estimate the noise and the signal according to our suggested methodology to fit the mixed interpolator $\dot{\beta}^I_{\hat{\alpha}_I}$. We also fit a semi-oracle estimator, $\dot{\beta}^I_{\hat{\alpha}_I(\tau)}$, which uses $\tau^2$ as if it were known and only estimates the noise with formula (\ref{F-Sigma}), and one oracle estimator, $\dot{\beta}_{\dot{\alpha}_I}$, using the oracle mixing ratios as if both $\tau^2$ and $\sigma^2$ were known.

\begin{figure}[ht!]  
\centering
\includegraphics[width=9cm]{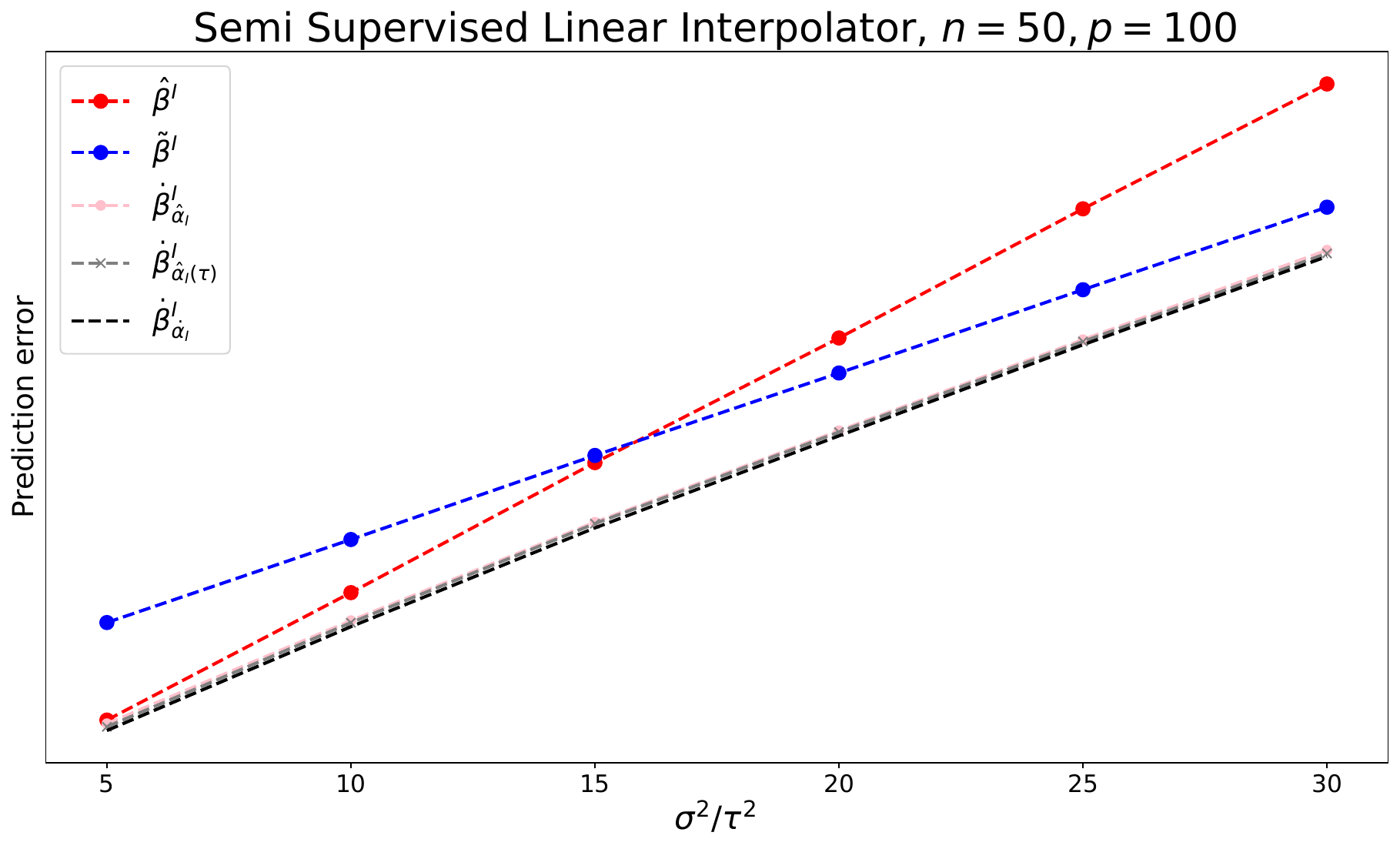}
\caption{Mean reducible errors of supervised and semi-supervised linear interpolating model, for various values of $\sigma^2$, with $\tau^2=1$, $n=50$ and $p=100$ (see text for details).}
\label{f4} 
\end{figure}

In the results (Figure \ref{f4}), we can see that $\dot{\beta}^I_{\hat{\alpha}_I}$ outperforms all non-mixed interpolators, suggesting that our mixing methodology is indeed useful in the over-parameterized regime. We also observe that knowing the true noise and signal provides no substantial benefit over estimating them. The differences are significant whenever the markers do not overlap.

\begin{figure}[ht!]  
\centering
\includegraphics[width=9cm]{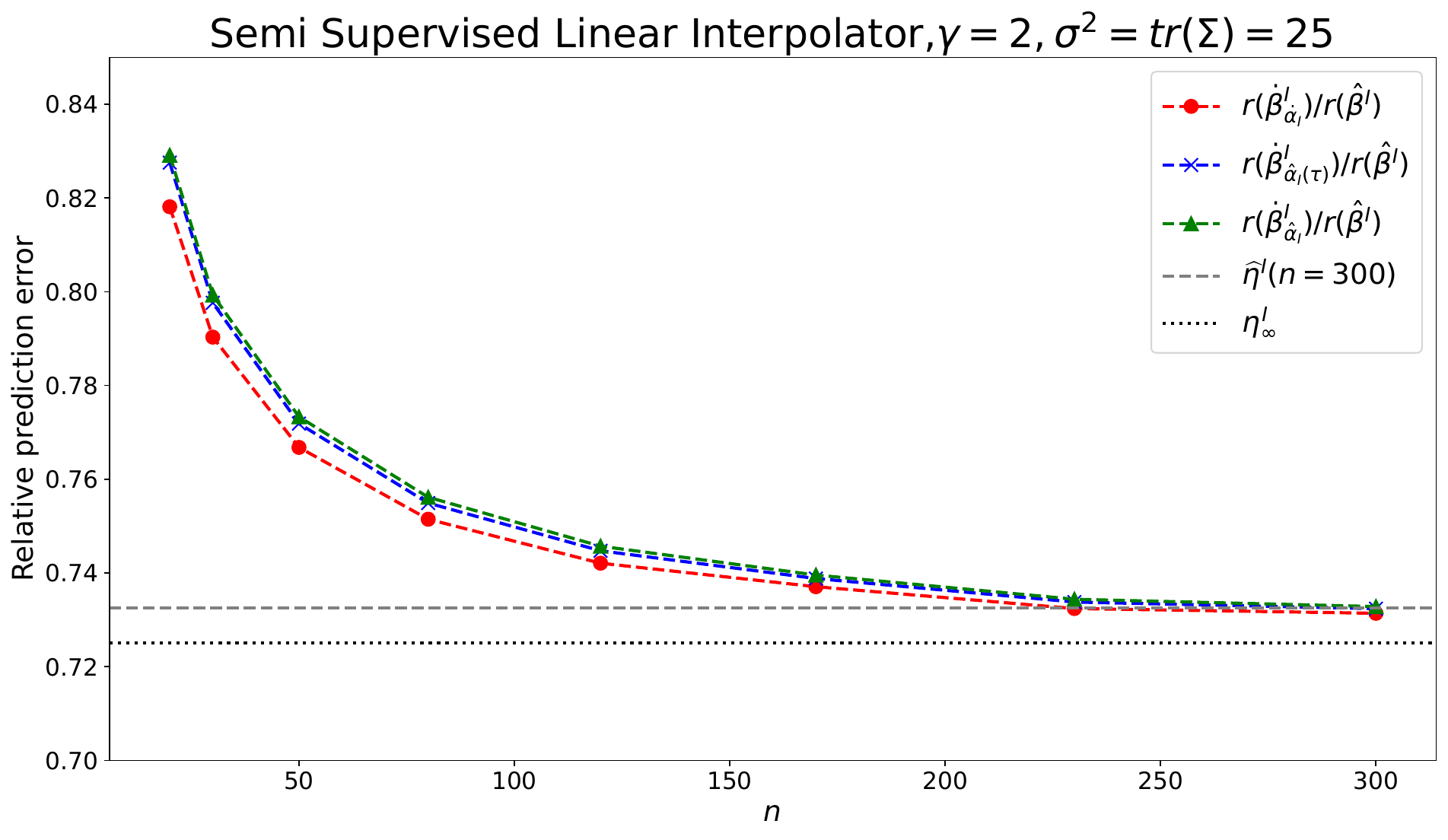}
\caption{Relative reducible errors of linear interpolating model for various values of $n$, with $\gamma=2$ and $\sigma^2=\text{tr}(\Sigma)=25$ (see text for details).}
\label{f5} 
\end{figure}

In Figure \ref{f5} we evaluate the predictive errors of $\dot{\beta}^I_{\hat{\alpha}_I}$, $\dot{\beta}^I_{\dot{\alpha}_I}$, $\dot{\beta}^I_{\hat{\alpha}_I(\tau)}$, and $\hat{\beta}^I$, for various multiple values of $n$ where $p=2n$, and with constants $\sigma^2=25$ and $\text{tr}(\Sigma)=25$, while maintaining the given structure for $\Sigma$ and incorporating a scaling factor. We then compare the subsequent empirical ratios: $r(\dot{\beta}^I_{\dot{\alpha}_I})/r(\hat{\beta}^I)$, $r(\dot{\beta}_{\hat{\alpha}_I(\tau)})/r(\hat{\beta}^I)$, and $r(\dot{\beta}^I_{\hat{\alpha}_I})/r(\hat{\beta}^I)$.
For reference, we indicate the theoretical boundary of $\eta^I$ in this setting with a dotted horizontal black line, and the estimated value of $\eta^I$ (based on unsupervised estimation with $n = 300$) with a dashed horizontal gray line. In the results, we can see that the relative predictive errors of all three estimators converge to the expected limiting value, supporting the theoretical analysis.

\section{Applications to deep learning - real data analysis}\label{Results-Real}
In this section we demonstrate the applicability of the suggested mixed-SSL to improve complex models such as deep neural networks (DNNs) with real data. In Section~\ref{LargeScale}, we investigate non-interpolating models applied to large image datasets, whereas Sections \ref{RFFSec} and \ref{FullyConnect} explore interpolating models using small tabular datasets of movie ratings. We propose intuitive modifications of the constructive methodology originally designed for GLM and linear interpolators to suit more complex models. Empirical results show that these proposed adaptations effectively decrease predictive error compared to standard supervised approaches. Throughout this section, the mean squared error (MSE) is referred to as the predictive error, denoted by $R$.

\subsection{Large-scale DNN - implementation on CelebA dataSet}\label{LargeScale}
In many state-of-the-art deep models, including models from the domain of image processing, the very last layer is linear with a relatively low dimension and uses the squared loss criterion. We suggest improving the predictive performance of these kinds of regression models by applying the mixed-loss-function methodology, which has been proven to be useful in GLM. The main idea is to view the last layer of the model as an isolated LM with random features, similar to the Conjugate Kernel method of \cite{hu2021random}. According to the results of the previous sections, this model can be improved by taking advantage of the unlabeled data, and a suitable mixing ratio can be found. The suggested method is described in the following algorithm.
\begin{enumerate}
    \item Train the network on the labeled data to achieve a supervised model $\hat{f}$.
    \item Feed forward the model $\hat{f}$ with the labeled and unlabeled data, \textbf{up to the last layer}, resulting in a new semi-supervised dataset $T^{\hat{f}}=(X^{\hat{f}},Y,Z^{\hat{f}})$.
    \item Fit a supervised linear predictor, $\hat{\beta}$, based on the training set $T^{\hat{f}}$ and estimate the noise and signal as follows:
\begin{align*}
    \hat{\tau}^2  = \hat{\beta}^t\Sigma\hat{\beta}/\text{tr}(\Sigma) \hspace{1mm};\hspace{1mm} \hat{\sigma}^2 =\widehat{Var}(Y)-\hat{\tau}^2\text{tr}(\Sigma),
\end{align*}
where $\Sigma$ is the empirical covariance matrix of $Z^{\hat{f}}$.
    \item Calculate an estimate $\hat{\alpha}$ of the desired mixing ratio, according to Eq. (\ref{E_alpha_dot}).
    \item Train the mixed model $\Ddot{f}_{\hat{\alpha}}$ on the original semi-supervised dataset $T=(X,Y,Z)$, starting from the model $\hat{f}$ and running a few more training epochs with the mixed loss function:
\[L^M_{\hat{\alpha}} = (1-\hat{\alpha})\hat{L}+\hat{\alpha} \Breve{L},\]
where $\hat{L}$ is the regular squared loss on the labeled data, and $\Breve{L}$ is equivalent to the semi-supervised loss in (\ref{L_breve_Def}), and can be written as follows:
\[\Breve{L} = \frac{1}{m}\sum_{i=1}^m \left(f(z_i)-\overline{Y}\right)^2 -2\widehat{\mathbb{C}ov}\left(f(X),Y\right).\]
\end{enumerate}

The described methodology has the advantage that the unlabeled data are used only in the last few epochs, thus saving computational resources during training. Moreover, it can be implemented in a stochastic manner by using mixed batches of labeled and unlabeled data.  We note that in Step $(5)$ the whole net's parameters are being updated during the training with the semi-supervised loss.

\begin{figure}[ht!]  
\centering
\includegraphics[width=9cm]{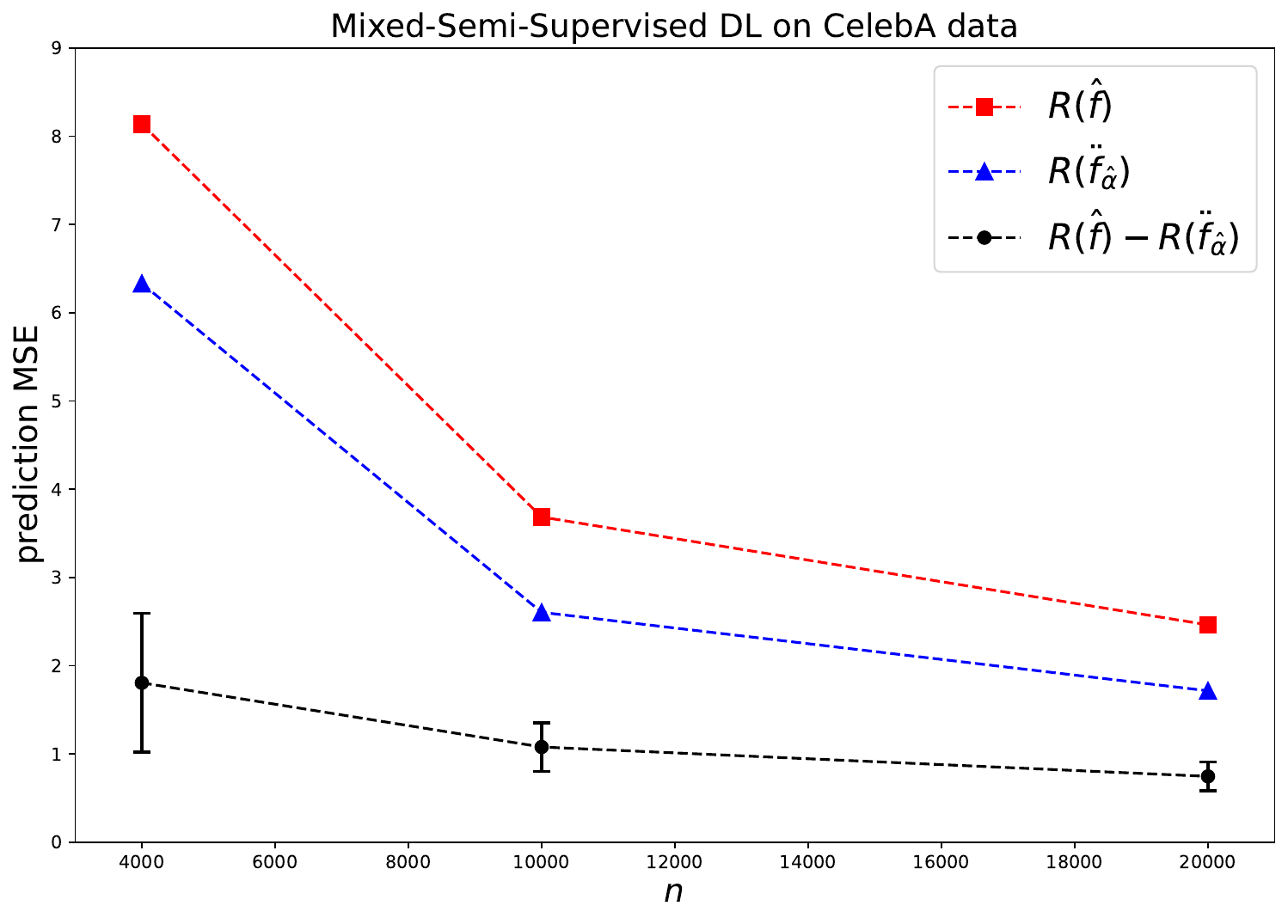}
\caption{Mean predictive errors of supervised and semi-supervised CNN on CelebA data-set, for various values of labeled training set size $n$, out of $2\cdot 10^5$ total images (see text for details).}
\label{CelebAFig}
\end{figure}

We demonstrate the above algorithm in the task of landmark localization in the CelebA facial images dataset (\cite{liu2018large}), containing $2\cdot 10^5$ images. We train the same Convolutional NN (CNN) as in \cite{simchoni2021using}, on a small fraction of the data designated as a labeled training set with additional $10$ semi-supervised epochs where the rest of the samples are used as an unlabeled set. The authors in  \cite{simchoni2021using} reported an MSE of \textbf{1.68} in the Nose-X prediction task with a training set of size $1.6\cdot 10^5$ labeled images. For comparison, in our experiment, with only $2\cdot 10^4$  labeled data, we obtain an MSE of \textbf{2.46} in the supervised learning and an MSE as low as \textbf{1.71} in the semi-supervised learning process. In order to achieve valid inference, we split the whole data into ten disjoint test folds while the unlabeled set is fixed and contains the whole dataset. For each test fold, the labeled training set is taken to be the following (or the first) $n$ images, where $n$ is the size of the training set. In figure \ref{CelebAFig}, we compare between the supervised model and the semi-supervised model, for different values of $n$. The majority of the hyper-parameters are fixed for any value of $n$, but the size of the batches and the number of supervised epochs vary with the training size $n$ to ensure the stability of the supervised learning. We present two learning curves of mean MSE, one for each learning method, and also one curve for the mean difference in MSE between the two methods, along with its $95\%$ confidence interval. In the results, we can see that our suggested SSL method delivers substantial and significant improvement over the standard supervised learning, and that the magnitude of the difference increases as $n$ decreases. 

\subsection{Application to Random Fourier Features on Netflix data}\label{RFFSec}
We demonstrate the applicability of the mixed-SSL interpolator on a common model in the domain of over-parametrized machine learning, called Random Fourier Features (RFF, \cite{rahimi2007random}). The RFF model can be viewed as a two-layer neural network, with the first layer being fixed (generated once pre-training), fully connected, with non-linear activation. The second (output) layer is essentially a linear interpolator. In particular, we adopt the following formulation for the output function of the RFF model:
\begin{align*}
    f(x)= \beta^t \varphi(Cx) \hspace{2mm};\hspace{2mm} C\in \mathbb{R}^{h\times p}  \hspace{1mm},\hspace{1mm} h>n  \hspace{2mm};\hspace{2mm} C_{ij}\sim N(0,1),
\end{align*}
where $\varphi$ is a non-linear function acting component-wise, and $w\in \mathbb{R}^h$ is some coefficients vector. Thus, we can fit supervised and semi-supervised interpolators, $\hat{\beta}^I$ and $\tilde{\beta}^I$, as well as the mixed interpolator $\dot{\beta}^I_{\alpha}$, based on the semi-supervised training set $\left(\varphi(XC^t),Y, \varphi(ZC^t)\right)$. 

We empirically study the predictive performance of the linear interpolators suggested in Section \ref{Interpolators-Sec-New} as well as the LM estimators from Section \ref{semi-supervised-GLM}, on a sub-dataset of the Netflix Prize data (\cite{bennett2007netflix}), available at \cite{netflix-prize-data}\footnote{https://www.kaggle.com/datasets/netflix-inc/netflix-prize-data}. From the full dataset, we extracted the $N=12,931$ users who rated the movie \say{Miss Congeniality}. The outcome $y$ is the rating of that movie by a specific user, and the covariate vector $x$ is the user's ratings of $p=40$ other movies. In order to achieve valid inference, we randomly split the data into $20$ disjoint folds, and each of them is split into train and test sets. This is done for different sizes of training sets $n$, while the unlabeled dataset $Z\in \mathbb{R}^{N\times p}$ is fixed and contains the rating data of all $N$ users, without the outcome column. For each of the training sets, we fit the LM estimators $(\hat{\beta},\Breve{\beta},\dot{\beta}_{\hat{\alpha}},\ddot{\beta}_{\tilde{\alpha}})$. We also generate an RFF model with $h=5n$ nodes in the hidden layer, and an activation function $\varphi:\mathbb{R}\to \mathbb{R}^3$ as follows: $\varphi(c) = \left(tanh(c), Sigmoid(c), \text{ELU}(c) \right)^t$, resulting in an output layer of size $15n$. We also apply standard scaling among $ \varphi(ZC^t)$, and then fit the interpolators $(\hat{\beta}^I,\tilde{\beta}^I,\dot{\beta}^I_{\hat{\alpha}_I})$. The mean squared error over the test set of the relevant fold is evaluated for every fitted estimator and averaged over the $20$ folds. The outcome is seven curves describing the predictive error, changing with $n$ for each of the estimators. We also evaluate the pairwise variances of the differences between two estimators, for any value of $n$, to find the minimum difference required for significance at any desired level.   

\begin{figure}[ht]  
\centering
\includegraphics[width=9cm]{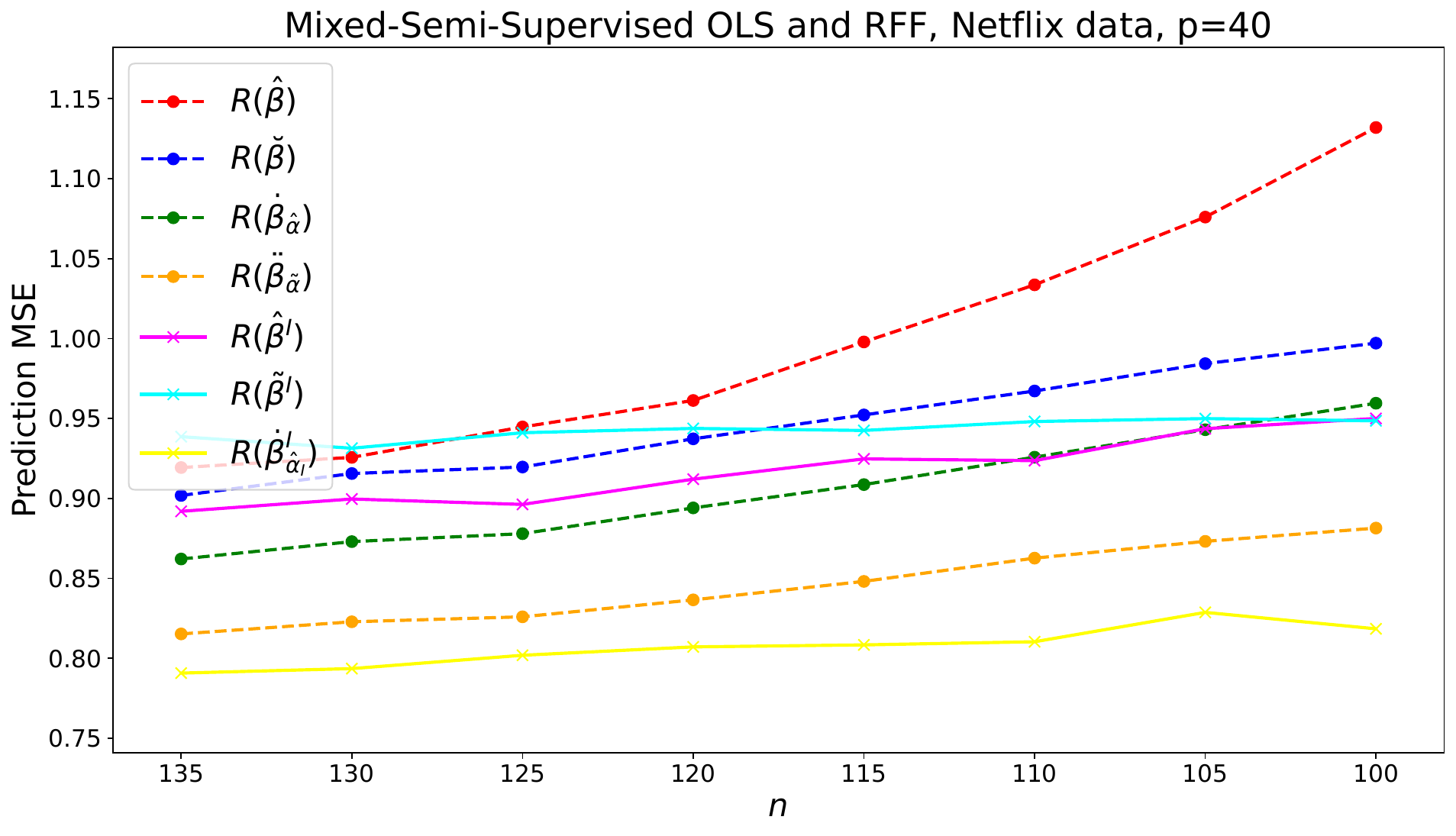}
\caption{Mean predictive errors of supervised and semi-supervised LM and RFF on Netflix data, for various values of $n$ (see text for details).}
\label{f6} 
\end{figure}

In the results (Figure \ref{f6}), we can see that the LM estimators behave with the same characteristics as in the simulated data of Section \ref{OLS_Simulations}, with $\ddot{\beta}_{\tilde{\alpha}}$ outperforming the others, supporting the analysis of Section \ref{semi-supervised-GLM}. Regarding the RFF model, we can see that compared to the LM, the RFF model is less sensitive to the size of the training set. More importantly, the mixed-RFF interpolator $\dot{\beta}^I_{\hat{\alpha}_I}$ is very useful in reducing predictive error and outperforms all other models. Any difference greater than $0.04$ is significant at the $0.05$ level.

\subsection{Application to Fully-Connected Interpolating Neural Networks}\label{FullyConnect}
We present a methodology for adapting the idea of mixed interpolators to fully connected deep neural networks. The main idea is to mix the predictions between two neural networks (models) as follows:
\begin{enumerate}
\item Supervised-trained interpolating model, denoted by $\hat{f}$, which is related to the minimum-norm interpolator.

\item  Semi-supervised-trained model, $\tilde{f}$, that aims to reduce the following variance term: $\mathbb{E}_{x_0} \left(\tilde{f}(x_0)-\overline{Y} \right)^2$, while approximately maintains the interpolation constraint over the labeled data.  
\end{enumerate}

The supervised model $\hat{f}$ can be trained by applying the standard gradient descent method on the labeled data, with square loss. The semi-supervised model, $\tilde{f}$, can be trained by solving the following unconstrained optimization problem:
\begin{align}\label{OptZ}\nonumber
    &\underset{\lambda \in \mathbb{R}^n}{\max} \hspace{1mm} \underset{w }{\min} \left\{ J(\lambda,w)   \right\}= \\ &  \underset{\lambda \in \mathbb{R}^n}{\max} \hspace{1mm} \underset{w }{\min} \left\{ \frac{1}{m}\sum_{i=1}^m \left(\tilde{f}(z_i)-\overline{Y}\right)^2-\lambda^t\left(Y-\tilde{f}(X)\right)   \right\}, 
\end{align}
where $w$ represents the model's parameters. Training $\tilde{f}$ can be done using a \say{simple} gradient descent ascent (GDA) method, minimizing $J$ w.r.t. $w$ and maximizing w.r.t. $\lambda$. However, the following issues need to be addressed in order to implement such an algorithm:
\begin{itemize}
    \item Starting point. We use the supervised interpolating model $\hat{f}$ as a starting point to train the model $\tilde{f}$.
    This is much more efficient than starting from a random point and converging to such an interpolating solution.   
    
    \item Stopping criterion. The suggested GDA method might \say{over-fit} the unlabeled data $Z$, when applied with a high-capacity model, resulting in poor predictions. Therefore, we define a minimal value for $\frac{1}{m}\sum_{i=1}^m \left(\tilde{f}(z_i)-\overline{Y}\right)^2$, where the GDA algorithm should be stopped when this value is reached.
\end{itemize}

The methodology suggested for the \textit{ mixed semi-supervised deep interpolation model} is described in the following algorithm.
\begin{enumerate}
    \item Train the desired neural network on the labeled data to achieve near-interpolating model $\hat{f}$.
    \item Feed forward the model $\hat{f}$ with the labeled and unlabeled data, \textbf{up to last layer}, and extract a new semi-supervised dataset $\mathcal{T}=(X^{\hat{f}},Y,Z^{\hat{f}})$ of high dimension $h>n$, with scaled covariates. 
    \item Based on $\mathcal{T}$, fit the supervised and semi-supervised linear interpolating models ($\hat{\beta}^I$ and $\tilde{\beta}^I$), and calculate an estimate of $\dot{\alpha}_I$ (denoted by $\hat{\alpha}_I$), according to formula (\ref{alpha_star_int}), combined with the following estimates of the noise and the signal:
\begin{align*}
    \hat{\tau}^2  = (\hat{\beta}^I)^t\Sigma\hat{\beta}^I/\text{tr}(\Sigma) \hspace{3mm};\hspace{3mm} 
   \hat{\sigma}^2 = \widehat{Var}(Y)-\hat{\tau}^2\text{tr}(\Sigma),
\end{align*}
where $\Sigma$ is the empirical covariance matrix of $Z^{\hat{f}}$.
    \item Train the semi-supervised model, $\tilde{f}$, by applying the GDA method with the objective $J$ (Equation~\ref{OptZ}), starting from $\hat{f}$, and converging to an interpolation model with a variance term as low as $(\tilde{\beta}^I)^t\Sigma \tilde{\beta}^I+\overline{Y}^2$.
    \item The mixed-semi-supervised deep model, denoted by $\dot{f}_{\hat{\alpha}_I}$, is defined follows:
    \begin{align*}
    \dot{f}_{\hat{\alpha}_I}(x) = \hat{\alpha}_I \tilde{f}(x)  +(1- \hat{\alpha}_I)\hat{f}(x) .
    \end{align*}
\end{enumerate}

We empirically study the predictive performance of $\dot{f}_{\hat{\alpha}}$ on the same Netflix data as in Section \ref{RFFSec}, for a smaller range of training sizes $n$. For any training sample, we train the model $\dot{f}_{\hat{\alpha}_I}$ as described above. We use a fully connected neural network with two hidden layers, containing $h_0=5n$ and $h=15n$ nodes, respectively, and ReLU activation. The mean square error over the unseen labeled in each fold is evaluated for both $\hat{f}$ and $\dot{f}_{\hat{\alpha}}$, and averaged over the $20$ folds. 

\begin{figure}[ht!]  
\centering
\includegraphics[width=9cm]{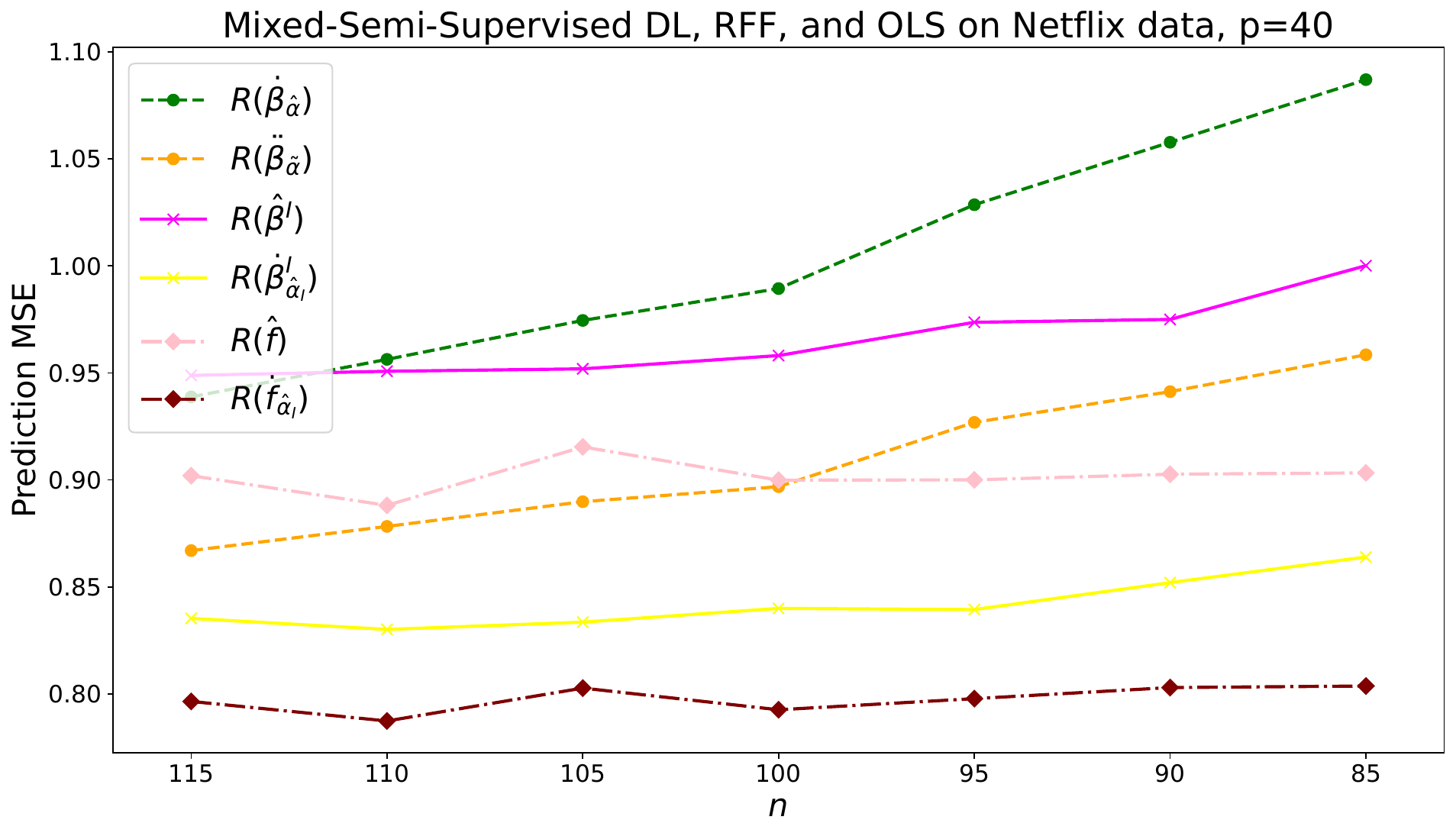}
\caption{Mean predictive errors of supervised and semi-supervised LM, RFF, and fully connected NN on Netflix data, for various values of $n$ (see text for details).}
\label{f7} 
\end{figure}

In Figure \ref{f7}, we present the results and compare them with the performance curves of the mixed LM and RFF models. We can see that the deep model is better than the other two models in both the supervised and semi-supervised settings. More importantly, the mixed-semi-supervised deep model, $\dot{f}_{\hat{\alpha}_I}$, improves over the supervised model $\hat{f}$, and delivers the best performance among all models. Any difference greater than $0.05$ is significant at the $0.05$ level.

\section{Discussion}\label{DiscussionSec}
This work provides new insight into the usefulness of unlabeled data in improving prediction models, with a focus on the GLM class. We show theoretically that unlabeled data can always be integrated into the learning process in a manner that ensures improvement in prediction. Furthermore, the optimal portion of integration or mixing (denoted by $\alpha$) can be well estimated in practice using the unlabeled data on hand, leading to substantial improvement. This kind of observation is very different from the message conveyed in many previous works about the usefulness of SSL.
Due to space limits, we study and analyze the methodology of mixed-SSL GLM and interpolators under a specific set of assumptions and conditions that we found most important and interesting. However, the suggested methodology can similarly be utilized in various situations, including diverse mechanisms for conditional variance, alternative predictive quality criteria for model evaluation, and different mixing methods. In essence, the broad concept of the methodology examined can be outlined as follows. 
\begin{enumerate}
\item Define the mixed predictor $f^M_{\alpha}$ that combines the unlabeled data during the learning process, with a portion determined by $\alpha$ and some mixing mechanism $M$.
\item Analyze the quantity of interest $R(f^M_{\alpha})$ according to the assumptions on the true model, and derive an approximate or heuristic formula of it, denoted by $r^M(\alpha)$.
\item Identify and calculate an estimation $\hat{\alpha}$ of the optimal mixing ration, where $r^M(\alpha)$ achieves its minimum, using estimates of the relevant statistics based on the labeled and unlabeled data.
\item Fit the model $f^M_{\hat{\alpha}}$.
\end{enumerate}

Based on the theoretical results in this paper, combined with the supporting empirical demonstrations, we believe that the general approach of mixed-SSL can be relevant for improving many real-life predictive models. However, the transition from GLM to more complex models is not a minor matter, especially in the context of obtaining rigorous theoretical insights. While this is not the main focus of this work, we find it important to demonstrate the applicability of our method to improve modern deep models in Section~\ref{Results-Real}. This was done by looking at the last layer of the model as a \say{remote} linear model with some best mixing ratio that can be estimated with the formulas in this paper. In our experiments, this estimation also turned out to be a useful heuristic measure for the desired mixing ratio of the whole deep model. We note that this heuristic approach was inspired by the fact that the estimator $\ddot{\beta}_{\hat{\alpha}}$ was very useful, although it used the best estimated mixing ratio of $\dot{\beta}_{\hat{\alpha}}$ as a heuristic measure. 

Nevertheless, more detailed theoretical analysis of the characteristics of mixed-SSDL is an important topic for future studies. In particular, it is interesting to study different adjustments of unlabeled data integration mechanisms to different classes of models, and identify how the performance changes with the portion of integration. We believe that both the theoretical and empirical results in this paper shed some light on this domain of research, and motivate follow-up work. Another promising direction for future research involves exploring the connection between the mixed-SSL strategy and modern self-supervised learning approaches. In the context of large language models (LLMs) and vision-language models (VLMs), many recent methods \citep{grill2020bootstrap,caron2021emerging,bao2021beit,oquab2023dinov2} aim to learn robust representations of inputs (whether text, images, or multimodal data) using large-scale unlabeled datasets. These representations are then adapted to downstream tasks using conventional models as adaptation heads, with a small amount of labeled data during a supervised fine-tuning phase. This setup shares strong similarities with the mixed-SSL framework presented in this paper. Based on our results, it is plausible that a mixed-SSL formulation could be used in the supervised fine-tuning stage to further leverage the unlabeled data.

\begin{appendix}
\section{Proofs}\label{appA}

\subsection{Proof of Theorem~\ref{P2n}} \label{appA1}
We begin by differentiating the function $\dot{r}_{\beta}(\alpha)$ in order to study its behavior at the endpoints and determine the minimizer. Computing the first and second derivatives yields
\[\frac{\partial}{\partial \alpha} \dot{r}_{\beta}(\alpha)  \mid _{\alpha=0}= -\sigma^2(v_l-v_u)    \hspace{2mm}; \hspace{2mm}    \frac{\partial}{\partial \alpha}\dot{r}_{\beta}(\alpha)   \mid _{\alpha=1}= 
    2\dot{B}_{\beta}(\alpha=1)> 0, \]
and
\[\frac{\partial^2}{\partial \alpha^2} \dot{r}_{\beta}(\alpha) = 2\dot{B}_{\beta}(\alpha=1)+\sigma^2(v_l-v_u).\]
These expressions imply that $\alpha=1$ cannot be the minimizer, as the first derivative is positive at that point. To further characterize the curvature, we analyze the difference $v_l-v_u$:
\begin{align*}
    v_l-v_u &= \frac{1}{n}\left[ \text{tr}\left( \mathbb{E}_X\left[ (X^tDX)^{-1}  \right]H \right)- \frac{n-1}{n}\text{tr}\left( H^{-1}  H \right)   \right] \\ & > \frac{1}{n} \text{tr}\left( \left(\mathbb{E}_X\left[ (X^tDX)^{-1}  \right] - H^{-1 }\right) H \right) \\ &=  \frac{1}{n} \text{tr}\left( \left(\mathbb{E}\left[ \left((D^{1/2}X)^tD^{1/2}X\right)^{-1}  \right] - \left(  \mathbb{E}\left[  (D^{1/2}X)^tD^{1/2}X  \right]   \right)^{-1 }\right) H \right).
\end{align*}
Under Assumption \ref{A0n}, the result of \cite{groves1969note} ensures that
\[\mathbb{E}\left[ \left((D^{1/2}X)^tD^{1/2}X\right)^{-1}  \right] - \left(  \mathbb{E}\left[  (D^{1/2}X)^tD^{1/2}X  \right]   \right)^{-1 }\] is positive semi-definite. Therefore, $v_l-v_u>0$, which in turn implies 
\begin{align*} \nonumber
    \frac{\partial}{\partial \alpha} \dot{r}_{\beta}(\alpha)  \mid _{\alpha=0} < 0   \hspace{2mm}; \hspace{2mm}   \frac{\partial^2}{\partial \alpha^2} \dot{r}_{\beta}(\alpha) > 0.
\end{align*}
These inequalities establish that $\alpha=0$ is not the minimizer and that $\dot{r}_{\beta}(\alpha)$ is strictly convex. Thus, a unique minimizer exists in the open interval $(0,1)$. 

To locate this minimizer, we set the first derivative to zero and solve for $\alpha$, obtaining
\begin{align*}
    \dot{\alpha}=\frac{\sigma^2(v_l-v_u) }{\dot{B}_{\beta}(\alpha=1)+\sigma^2(v_l-v_u)}.
\end{align*}
Finally, substituting $\dot{\alpha}$ into the expressions for $\dot{B}_{\beta}(\alpha)$ and $\dot{V}_{\beta}(\alpha)$, we find the explicit form of the minimized value:
\begin{align*}
    \dot{r}_{\beta}(\alpha)  &= \frac{1}{2}\left[\sigma^2v_l- \frac{\sigma^4(v_l-v_u)^2 }{\dot{B}_{\beta}(\alpha=1)+\sigma^2(v_l-v_u) }\right] \\ &=  \dot{r}_{\beta}(\alpha=0) -  \frac{1}{2} \cdot\frac{\sigma^4(v_l-v_u)^2 }{\dot{B}_{\beta}(\alpha=1)+\sigma^2(v_l-v_u) } < \dot{r}_{\beta}(\alpha=0) & \qed
\end{align*}

\subsection{Proof of Theorem~\ref{P4n}}\label{appA4n}
We begin by analyzing the variance component of $\dot{r}(\alpha)$, which can be expressed as follows:
\begin{align}\label{Vdot_Term1}
    \dot{V}_{\beta}(\alpha)&=\frac{\sigma^2}{2}\left[ (1-\alpha)^2 v_l +\alpha^2v_u +2\alpha(1-\alpha)v_u  \right] \\&=\frac{\sigma^2}{2n}\text{tr}\left(H\mathbb{E}_{X}\left[ \tilde{S}_{\alpha}  X^tDX \tilde{S}_{\alpha}\right]\right) - \frac{\sigma^2}{2(n-1)}\left(\alpha^2 v_u +2\alpha(1-\alpha )v_u \right)\label{Vdot_Term2}
\end{align}
where $\tilde{S}_{\alpha}=\alpha H^{-1} + (1-\alpha)(X^tDX)^{-1}$. Since it is known that $v_l>v_u$, we can establish the following inequality:
\[\frac{\alpha^2 v_u +2\alpha(1-\alpha )v_u }{n-1} \leq \frac{1-\xi_{\alpha}}{n}\text{tr}\left(H\mathbb{E}_{X}\left[ \tilde{S}_{\alpha}  X^tDX \tilde{S}_{\alpha}\right]\right),\]
and together with (\ref{Vdot_Term2}), we conclude that:
\begin{align*}
    \dot{V}_{\beta}(\alpha)& \geq \frac{ \xi_{\alpha}\sigma^2}{2n}\text{tr}\left(H\mathbb{E}_{X}\left[ \tilde{S}_{\alpha}  X^tDX \tilde{S}_{\alpha}\right]\right).
\end{align*}
Next, we establish the following identity, which holds for any value of $\alpha$ and for any realization of $X$:
\begin{align*}
    \tilde{S}_{\alpha}  X^tDX S_{\alpha} = & \hspace{1mm}\Big(\alpha^2H(X^tDX)^{-1}H+(1-\alpha)^2 X^tDX(X^tDX)^{-1}X^tDX \\ &+\alpha(1-\alpha)H(X^tDX)^{-1}X^tDX \\&+\alpha(1-\alpha)X^tDX(X^tDX)^{-1}H\Big)^{-1} \\
    =& \Big(\alpha^2H(X^tDX)^{-1}H+(1-\alpha)^2X^tDX  +2\alpha(1-\alpha)H \Big)^{-1}
    \\&    =  S_{\alpha}  X^tX \tilde{S}_{\alpha}.
\end{align*}
With this identity in hand, we are able to directly compare the two variance terms, as demonstrated below:
\begin{align*}
    \dot{V}_{\beta}(\alpha)-\Ddot{V}_{\beta}(\alpha) & \geq \frac{ \xi_{\alpha}\sigma^2}{2n}\text{tr}\left(H\mathbb{E}_{X}\left[ \tilde{S}_{\alpha}  X^tDX \tilde{S}_{\alpha} - S_{\alpha}  X^tDX S_{\alpha} \right]\right) \\ &= \frac{ \xi_{\alpha}\sigma^2}{2n}\mathbb{E}_{X}\left[\text{tr}\left(H \left(\tilde{S}_{\alpha}-S_{\alpha}\right) X^tDX \left(\tilde{S}_{\alpha}+S_{\alpha}\right) \right)\right].
\end{align*}
In the expression above, we observe that $\tilde{S}_{\alpha}-S_{\alpha}$ is a positive definite matrix, due to the convexity of the inverse operator, combined with the fact that both $H$ and $X^tDX$ are positive definite matrices, as stated by \cite{nordstrom2011convexity}. Furthermore, the matrix $ \left( \tilde{S}_{\alpha} - S_{\alpha} \right)X^tDX\left( \tilde{S}_{\alpha} + S_{\alpha} \right)$ is semi-positive definite because it is composed of a product of positive definite matrices and is symmetric (it can be expressed as the difference between two symmetric matrices). We conclude that the term inside the trace is a product of two semi-positive definite matrices. Hence, the trace is non-negative for every realization of X, and its expectation is also non-negative. Therefore, for any given mixing ratio $\alpha$, we obtain the inequality:
\[ \dot{V}_{\beta}(\alpha) \geq \Ddot{V}_{\beta}(\alpha).\]
On the other hand, from equation (\ref{Vdot_Term1}) we find that:
\[\frac{\partial}{\partial \alpha}\dot{V}_{\beta}(\alpha) \mid _{\alpha=0} = \sigma^2(v_u-v_l)<0.\]
Since $\dot{V}_{\beta}(\alpha=0)=\Ddot{V}_{\beta}(\alpha=0)$, together with the fact that $\Ddot{V}_{\beta}(\alpha) \leq \dot{V}_{\beta}(\alpha)$, we conclude that necessarily $\frac{\partial}{\partial \alpha}\Ddot{V}_{\beta}(\alpha) \mid _{\alpha=0} <0$.

By analyzing the first derivative of $\ddot{V}_{\beta}(\alpha)$ at $\alpha=1$, we can show the following:
\begin{align*}
\frac{\partial}{\partial \alpha}\ddot{V}_{\beta}(\alpha) \mid _{\alpha=1} &= \frac{\sigma^2(n-1)}{n^2} \text{tr}\left(\mathbb{E}_{X}\left[H^{-1}X^tDXH^{-1}X^tDX\right]  -H^{-1}X^tDX  \right)
\\&= \frac{\sigma^2(n-1)}{n^2} \text{tr}\left(\mathbb{E}_{X}\left[(H^{-1/2}X^tDXH^{-1/2})^{2}\right]  -I_p  \right)\\ &
> \frac{\sigma^2(n-1)}{n^2} \text{tr}\left(\left(\mathbb{E}_{X}\left[H^{-1} X^tDX\right]\right)^{2}  -I_p  \right) =  0.
\end{align*}
In the above derivations, we employ the general findings from \cite{rahman1973note}. This implies that for a symmetric, positive definite random matrix $A$ and any integer $k$, the matrix $\mathbb{E}\left[A^{k} \right] - \left(\mathbb{E}[A]\right)^{k}$ is positive semi-definite provided that $A$ is not deterministic. Specifically, we apply the observation that $\text{tr}\left(\mathbb{E}\left[A^{2} \right] \right)> \text{tr}\left( \left(\mathbb{E}[A]\right)^{2}\right)$ and choose $A = H^{-1/2}X^tDXH^{-1/2}$.

By analyzing the bias component of $\Ddot{r}_{\beta}(\alpha)$, we can immediately find that:
\[\frac{\partial}{\partial \alpha}\Ddot{B}_{\beta}(\alpha) \mid _{\alpha=0}  = 0.\]
Now, let us look at the following expression of the first derivative at $\alpha=1$:
\begin{align*}
\frac{\partial}{\partial \alpha}& \Ddot{B}_{\beta}(\alpha) \mid _{\alpha=1} =  \frac{1}{n}\mathbb{E}_{X}\left[\text{tr}\left(H^{-1}X^tDX H^{-1} \zeta \zeta^t\right)\right] .
\end{align*}
The expression above is necessarily non-negative, since we have a product of two positive semi-definite matrices inside the trace. From all the above, we conclude that:
\begin{align*}
     \frac{\partial \Ddot{r}_{\beta}(\alpha)}{\partial \alpha}  \mid _{\alpha=0} &=\frac{\partial  \left[ \Ddot{B}_{\beta}(\alpha) +  \Ddot{V}_{\beta}(\alpha)\right]}{\partial \alpha}  \mid _{\alpha=0} < 0   \\   \frac{\partial \Ddot{r}_{\beta}(\alpha)}{\partial \alpha}  \mid _{\alpha=1} &=\frac{\partial  \left[ \Ddot{B}_{\beta}(\alpha) +  \Ddot{V}_{\beta}(\alpha)\right]}{\partial \alpha}  \mid _{\alpha=1}>0.  
\end{align*}
Thus, we conclude that $\arg\min\{\Ddot{r}_{\beta}(\alpha)\}$ is in the open interval $(0,1)$. \qed

\subsection{Proof of Theorem~\ref{P2.11n}}\label{appA4}
Within the linear framework, the relationships $B_{\beta}(\ddot{\beta}_{\alpha})=\ddot{B}_{\beta}(\alpha)$ and $B_{\beta}(\dot{\beta}_{\alpha})=\dot{B}_{\beta}(\alpha)$ hold, and we consistently apply the given expressions for $\ddot{B}_{\beta}(\alpha)$ and $\dot{B}_{\beta}(\alpha)$ throughout this proof. We express the approximate function $\Ddot{B}_{\beta}(\alpha)$ in the following way:
\[ \ddot{B}_{\beta}(\alpha) \approx \Ddot{b}(\alpha) = b_0+b_1\alpha+b_2\alpha^2+b_3\alpha^3. \]
In general, $ \ddot{B}_{\beta}(\alpha)$ satisfies the following:
\[ \ddot{B}_{\beta}(\alpha) \mid _{\alpha=0}= \frac{\partial}{\partial \alpha}  \ddot{B}_{\beta}(\alpha) \mid _{\alpha=0} =0,\]
\[ \ddot{B}_{\beta}(\alpha) \mid _{\alpha=1}=   \dot{B}_{\beta}(\alpha) \mid _{\alpha=1} = \frac{1}{2n} \text{tr}\left(H^{-1}  \mathbb{E}_{X}\left[ \zeta \zeta^t \right]   \right) := \dot{B}_1 .\]
By restricting $\Ddot{b}(\alpha)$ to satisfy the  conditions
\[\Ddot{b}(0)=  \ddot{B}_{\beta}(0)   \hspace{1mm};\hspace{1mm}  \ddot{b}'(0)= \frac{\partial \ddot{B}_{\beta}(\alpha) }{\partial \alpha} \mid _{\alpha=0},\] we get $b_0=b_1=0$. By restricting $\Ddot{b}(\alpha)$ to satisfy $\Ddot{b}(1)= \ddot{B}_{\beta}(1) $, we get $b_2+b_3= \dot{B}_1$. By the definition of $\dot{B}_{\beta}(\alpha)$, we can see that for any $\alpha$, $\dot{B}_{\beta}(\alpha)=\alpha^2 \dot{B}_1$, and therefore:
\[\frac{\partial}{\partial \alpha} \dot{B}_{\beta}(\alpha)) \mid _{\alpha=1}=2\dot{B}_1.\]

In the specific case of a linear model, $\mu = X\beta$, $H=\mathbb{E}[X^tX]$ , and the terms $\alpha \zeta$ and $\alpha S_{\alpha}\zeta$ can be simplified and written as follows:
\begin{align*}
    \alpha \zeta&= \alpha \left( H+ n\overline{X}\cdot \overline{X}^t - X^tX \right)  \beta, \\
    \alpha S_{\alpha}\zeta&= \alpha S_{\alpha} \left( H+ n\overline{X}\cdot \overline{X}^t - X^tX \right)  \beta \\&= 
    S_{\alpha}\left( S_{\alpha}^{-1} +\alpha n\overline{X}\cdot \overline{X}^t -X^tX\right) \beta\\ &=
    \left(I-S_{\alpha} (X^tX-\alpha n\overline{X} \hspace{1mm} \overline{X}^t )\right)\beta,
\end{align*}
where $S_{\alpha}= \left( \alpha H + (1-\alpha)X^tX  \right)^{-1}$. Thus, the bias components $\ddot{B}_{\beta}(\alpha)$ and $\dot{B}_{\beta}(\alpha)$ can be written as follows:
\begin{align*}
\dot{B}_{\beta}(\alpha) &= \frac{\alpha^2}{2n} \text{tr}\left(H^{-1}\mathbb{E}_X\left[\zeta \zeta^t  \right] \right) \\ &= \frac{\alpha^2}{2n}\text{tr}\left(\mathbb{E}\left[ \beta \beta^t \left( H+ n\overline{X}\cdot \overline{X}^t - X^tX \right) H ^{-1}\left( H+ n\overline{X}\cdot \overline{X}^t - X^tX \right)  \right]\right),
    \\ \ddot{B}_{\beta}(\alpha) &= \frac{\alpha^2}{2n} \text{tr}\left(H\mathbb{E}_X\left[S_{\alpha}\zeta \zeta^t S_{\alpha} \right] \right) = \frac{1}{2n}\text{tr}\left(\mathbb{E}_{X}\left[ \beta \beta^t \Delta_{\alpha}^tH \Delta_{\alpha} \right]\right),
\end{align*}
where, $ \Delta_{\alpha} =S_{\alpha} (X^tX-\alpha n\overline{X} \hspace{1mm} \overline{X}^t ) -I$. In the derivations that follow, we use the following observation:
\begin{align}\label{CxObs}
    n\overline{X} \hspace{1mm} \overline{X}^t = \frac{1}{n} X^tX + C_X \hspace{1mm} ; \hspace{1mm} \left[ C_X\right]_{jk}= \frac{1}{n} \sum_{i\neq l } x_{ij}x_{lk}, \\ \nonumber \mathbb{E}\left[ C_X\right] = \mathbb{E}\left[ X^tX C_X\right] =\mathbb{E}\left[ (X^tX)^{-1} C_X\right]=\textbf{0}_{p\times p}.
\end{align}
We note that the term $C_X$ cancels out in many of the expectations that follow, unless it is squared.

Using the above, we can write the explicit expressions for the first derivatives of $\dot{B}_{\beta}(\alpha)$ and $ \ddot{B}_{\beta}(\alpha)$ at $\alpha=1$, specifically for the linear model, up to a factor of $((n-1)/n)^2$, as follows.
\begin{align*}
    \frac{\partial \dot{B}_{\beta}(\alpha)}{\partial \alpha}  \mid _{\alpha=1} &= \frac{2}{n}\text{tr}\left(\beta \beta^t\mathbb{E}_{X}\left[ X^tXH^{-1}X^tX-H+ C_XH^{-1}C_X \right]\right) \\
    \frac{\partial\ddot{B}_{\beta}(\alpha)}{\partial \alpha}  \mid _{\alpha=1} &=\frac{2}{n}\text{tr}\left(\beta \beta^t\mathbb{E}_{X}\left[ \left(X^tXH^{-1}\right)^2X^tX-2X^tXH^{-1}X^tX+H \right]\right) \\ &+\frac{2}{n}\text{tr}\left(\beta \beta^t\mathbb{E}_{X}\left[ C_XH^{-1}C_X \right]\right),
\end{align*}
where $C_X$ is defined in Equation (\ref{CxObs}).

In order to analyze the difference between the values of the derivatives at $\alpha=1$, we consider Assumption~\ref{AMechanisim_n}, and denote by $\tilde{X}\in \mathbb{R}^{n \times p}$ the random matrix with rows $\{\tilde{x}_i\}_{i=1}^n$, such that $X=\Tilde{X}\Sigma^{1/2}$, and the following identities hold:
\begin{align*}
    X^tX = \Sigma^{1/2}\tilde{X}^t\tilde{X}\Sigma^{1/2} \hspace{2mm}&;\hspace{2mm} \mathbb{E}\left[ \tilde{X}^t\tilde{X}\right] = nI_p\hspace{2mm};\hspace{2mm} H = n\Sigma,  \\ 
    \mathbb{E}\left[ (\tilde{X}^t\tilde{X})^2\right] &= n\left(n+p+q_4-2 \right)I_p, \\
    \mathbb{E}\left[ (\tilde{X}^t\tilde{X})^3\right] &=I_pn\Big(q_6 +3(p-1)q_4+p(p-3) \\ &+3(n-1)(q_4+p-1) + n(n-3) +4 \Big).
\end{align*}
Using the above properties, we can show that the following holds:
\begin{align*}
    \frac{\partial \ddot{B}_{\beta}(\alpha)}{\partial \alpha}  - \frac{\partial \dot{B}_{\beta}(\alpha)}{\partial \alpha}  \mid _{\alpha=1} &= 2   \beta^t\left(  \Sigma \mathbb{E}_{X}\left[ \frac{1}{n^3}(\widetilde{X}^t\widetilde{X})^3 -  \frac{3}{n^2}(\widetilde{X}^t\widetilde{X})^2  \right] +2\Sigma \right)\beta\\
    &= 2\beta^t\Sigma \beta\left[ \frac{3}{n}+ \frac{p(p-6)+3q_4(p-2)+7+q_6}{n^2}\right] >0.
\end{align*}
From the above, we conclude that the first derivative of $\ddot{B}_{\beta}(\alpha)$ at $\alpha=1$ is greater than the one of $\dot{B}_{\beta}(\alpha)$. We plug this result into the constraint $\ddot{b}'(1) = \frac{\partial \ddot{B}_{\beta}(\alpha) }{\partial \alpha}  \mid _{\alpha=1}$,  and we get that:
\begin{align*}
    2b_2+3b_3&>2\dot{B}_1\\
    2b_2+3(B_1-b_2)&>2\dot{B}_1\\
    \dot{B}_1&>b_2.
\end{align*}
Therefore, we can write:
\begin{align*}
    B_{\beta}(\ddot{\beta}_{\alpha}) \approx\Ddot{b}(\alpha) &= b_2\alpha^2 +\left(\dot{B}_1-b_2\right)\alpha^3 = b_2(\alpha^2-\alpha^3) + \dot{B}_1\alpha^3 &\\
    & \leq  \dot{B}_1(\alpha^2-\alpha^3+\alpha^3)= \alpha^2 \dot{B}_1= \dot{B}_{\beta}(\alpha) =B_{\beta}(\dot{\beta}_{\alpha}) &\qed
\end{align*}

\subsection{Proof of Proposition~\ref{P5n}}\label{appA110}
In the case of the linear model, with random $\beta$ mechanism, the reducible error for any estimator $\beta_T$ can be expressed as follows:
\begin{align*}
    r(\beta_T) =& \hspace{1mm} \underbrace{\frac{1}{2} \mathbb{E}_{X,x_0,\beta}\left[ \left( x_0^t\mathbb{E}[\beta_T \mid X,\beta] - x_0^t\beta   \right)^2 \right]   }_\text{$B(\beta_T)$} +\underbrace{\frac{1}{2}  \mathbb{E}_{X,x_0,\beta}\left[ \mathbb{V}ar\left( x_0^t\beta_T \mid X,x_0,\beta\right)\right] 
  }_\text{$V(\beta_T)$}.
\end{align*}
Focusing on the bias term, we can write:
\[B(\beta_T)=\frac{1}{2} \text{tr}\left(\Sigma\mathbb{E}_{X,\beta}\left[ \left( \mathbb{E}[\beta_T \mid X,\beta] - \beta   \right)\left( \mathbb{E}[\beta_T \mid X,\beta] - \beta   \right)^t \right]\right). \]
By taking
\[\beta_T= \dot{\beta}_{\alpha}^I = \alpha\tilde{\beta}^I+(1-\alpha)\hat{\beta}^I=S^I_{\alpha}Y,\]
where
\[S^I_{\alpha}= \alpha\Sigma^{-1}X^t(X\Sigma^{-1}X^t)^{-1}+(1-\alpha)X^t(XX^t)^{-1},\]
and using the fact that $\mathbb{E}[Y \mid X,\beta]=X\beta$, we obtain the following derivation:
\begin{align*}
   B(\dot{\beta}_{\alpha}^I) & =\frac{1}{2} \text{tr}\left(\Sigma\mathbb{E}_{X,\beta}\left[ \left( S_{\alpha}^IX\beta - \beta   \right)\left( S_{\alpha}^IX\beta - \beta   \right)^t \right]\right) \\ 
   &= \frac{1}{2} \text{tr}\left(\Sigma\mathbb{E}_{X,\beta}\left[ \left( S_{\alpha}^IX - I_p   \right)\beta \beta^t \left( S_{\alpha}^IX - I_p  \right)^t \right]\right). 
\end{align*}
Using the fact that the random mechanism of $\beta$ implies that $\mathbb{E}[\beta \beta^t]=I_p\tau^2$, we can simplify as follows:
\begin{align*}
   B(\dot{\beta}_{\alpha}^I) & =\frac{\tau^2}{2}  \text{tr}\left(\Sigma\mathbb{E}_{X,\beta}\left[ \left( S_{\alpha}^IX - I_p   \right) \left( S_{\alpha}^IX - I_p  \right)^t \right]\right)\\ &= \frac{\tau^2}{2} \left[ \alpha^2b_l^I+(1-\alpha)^2b_u^I+2\alpha(1-\alpha)b_l^I -2(\alpha b_l^I+(1-\alpha)b_u^I+\text{tr}(\Sigma)\right] \\ &= \frac{\tau^2}{2} \left[ \alpha^2(b_u^I-b_l^I)-b_u^I+\text{tr}(\Sigma)\right] . 
\end{align*}
Focusing on the variance term, and employing the same notations, we can write:
\begin{align*}
   V(\dot{\beta}_{\alpha}^I) & =\frac{1}{2} \text{tr}\left(\Sigma\mathbb{E}_{X,\beta}\left[  S_{\alpha}^I  \left(\mathbb{V}ar\left(Y \mid X,\beta\right)   \right)\left( S_{\alpha}^I\right)^t \right]\right),
\end{align*}
and using the fact that $\mathbb{V}ar(Y\mid X\beta)=I_{n}\sigma^2$, we obtain the following:
\begin{align*}
   V(\dot{\beta}_{\alpha}^I)  &=\frac{\sigma^2}{2} \text{tr}\left(\Sigma \mathbb{E}_{X}\left[  S^I_{\alpha} (S^I_{\alpha})^t \right] \right)  \\ 
   &= \frac{\sigma^2}{2} \left(\alpha^2 v_u^I +(1-\alpha)^2v_l^I  +2\alpha(1-\alpha) v_u^I   \right).  
\end{align*}

Concerning the relation between $b^I_l, b^I_u$ and between $v^I_l, v^I_u$, we note that by definition, for any given realization of $(\beta,X,Y)$:
\begin{align*}
    (\tilde{\beta}^I)^t\Sigma  \tilde{\beta}^I \leq  (\hat{\beta}^I)^t\Sigma  \hat{\beta}^I.
\end{align*}
Therefore, for any setting of $(P_x, \sigma^2, \tau^2)$, the integrated expectations hold that:
\begin{align*}
    \mathbb{E}\left[ (\tilde{\beta}^I)^t\Sigma  \tilde{\beta}^I\right] \leq  \mathbb{E}\left[ \hat{\beta}^I)^t\Sigma  \hat{\beta}^I \right].
\end{align*}
The derivation of the above expectations shows that:
\begin{align*}
 \mathbb{E}\left[ (\tilde{\beta}^I)^t\Sigma  \tilde{\beta}^I\right]&=\text{tr}\left(\mathbb{E}_{X}\left[ X^t(X\Sigma^{-1}X^t)^{-1}  \mathbb{E}\left[YY^t \mid X\right] (X\Sigma^{-1}X^t)^{-1}  X\Sigma^{-1}\right]\right)= \sigma^2v_u^I+\tau^2b_l^I,\\
 \mathbb{E}\left[ (\hat{\beta}^I)^t\Sigma  \hat{\beta}^I\right]&=\text{tr}\left(\Sigma \mathbb{E}_{X}\left[ X^t(XX^t)^{-1}  \mathbb{E}\left[YY^t \mid X\right] (XX^t)^{-1}  X\right]\right)= \sigma^2v_l^I+\tau^2b_u^I,
\end{align*}
and therefore we have:
\begin{align*}
   \tau^2 b_l^I+\sigma^2 v^I_u &\leq \tau^2 b_u^I+\sigma^2 v^I_l,
\end{align*}
Since the above inequality is valid for any configuration of $(\sigma^2, \tau^2)$, it follows logically that $b^I_l\leq b^I_u$ and $v^I_u\leq v^I_l$.
\qed

\subsection{Proof of Theorem~\ref{Interpol1}}\label{appA11}
The first and second derivatives of $r(\dot{\beta}^I_{\alpha}) $ with respect to $\alpha$ can be written as follows:
\begin{align*}
    \frac{\partial}{\partial \alpha}&r(\dot{\beta}^I_{\alpha})  = \alpha\left[  \tau^2(b^I_u-b^I_l)+\sigma^2(v^I_l-v^I_u)   \right] -\sigma^2(v^I_l-v^I_u), \\
    \frac{\partial}{\partial \alpha}&r(\dot{\beta}^I_{\alpha})   \mid _{\alpha=0} = -\sigma^2(v^I_l-v^I_u) < 0,  \hspace{2mm}; \hspace{2mm}  \frac{\partial}{\partial \alpha} r(\dot{\beta}^I_{\alpha})   \mid _{\alpha=1} = \tau^2(b^I_u-b^I_l)> 0, \\
    \frac{\partial^2}{\partial \alpha^2}&r(\dot{\beta}^I_{\alpha}) = \tau^2(b^I_u-b^I_l)+\sigma^2(v^I_l-v^I_u) > 0.
\end{align*}
From the above, we conclude that for any combination $(\tau^2,\sigma^2)$, it is beneficial to slightly move from each one of the estimators toward the other. Moreover, the best mixed interpolator is achieved where the first derivative equals $0$ as follows:
\begin{align}
     \dot{\alpha}_I=\frac{\sigma^2(v^I_l-v^I_u) }{\tau^2(b^I_u-b^I_l)+\sigma^2(v^I_l-v^I_u)}.
\end{align}
By substituting the above expression of $\dot{\alpha}_I$ into $r(\dot{\beta}_{\alpha})$, we get that:
\[r(\dot{\beta}^I_{\dot{\alpha}_I})= \frac{1}{2}\left[ \tau^2(\mathrm{tr}(\Sigma)-b_u)+\sigma^2 v^I_l -\frac{\sigma^4(v^I_l-v^I_u)^2 }{\tau^2(b^I_u-b^I_l)+\sigma^2(v^I_l-v^I_u)} \right],\]
and the reducible error of $\hat{\beta}^I$ is simply
\[r(\hat{\beta}^I)=\frac{1}{2}\left[ \tau^2(\mathrm{tr}(\Sigma)-b_u)+\sigma^2 v^I_l \right].\]
Thus, we conclude that:
\begin{align*}
    r(\dot{\beta}^I_{\dot{\alpha}_I})= r(\hat{\beta}^I)- \frac{1}{2}\left[\frac{\sigma^4(v^I_l-v^I_u)^2 }{\tau^2(b^I_u-b^I_l)+\sigma^2(v^I_l-v^I_u)} \right] < r(\hat{\beta}^I). \qed
\end{align*}
\subsection{Proof of Proposition~\ref{P6}}\label{appA12}
By taking iterative expectation over $X$, $\beta$, and $Y$, we get:
\[\mathbb{E}[\hat{\sigma}^2]=  \mathbb{E}_X \left[\frac{\mathbb{E}_{\beta}  \left[ \text{tr}\left( (XX^t)^{-2} \mathbb{E}_Y\left[  YY^t  \mid X,\beta \right]  \right)  \right] -\tau^2\text{tr}\left( (XX^t)^{-1} \right)  }{\text{tr}\left( (XX^t)^{-2}\right) }\right],\]
and since the underlying model satisfies $\mathbb{E}_Y\left[  YY^t  \mid X,\beta \right]=I_n \sigma^2+X\beta\beta^tX^t$, we can write:
\[\mathbb{E}[\hat{\sigma}^2] = \mathbb{E}_X \left[\frac{\text{tr}\left( (XX^t)^{-2} \sigma^2+X^t(XX^t)^{-2}X \mathbb{E}_{\beta}\left[\beta \beta^t\right]-  \tau^2(XX^t)^{-1}\right)}{\text{tr}\left( (XX^t)^{-2}\right) } \right].\]
Now, since $\mathbb{E}_{\beta}\left[\beta \beta  ^t\right]=I_p\tau^2$, the middle and right terms in the numerator cancel, and we are left with:
\begin{align*}
    &&\mathbb{E}[\hat{\sigma}^2]&=  \mathbb{E}_X\left[\frac{\text{tr}\left( (XX^t)^{-2}\right) \sigma^2}{\text{tr}\left( (XX^t)^{-2}\right) }    \right] = \sigma^2. && && \qed
\end{align*}

\subsection{Proof of Proposition~\ref{GassianAssym1}}\label{appA5}
    By the definitions of $\eta$ and $\dot{\alpha}$, their limiting behavior can be expressed by the limiting behavior of $v_l$, $v_u$, and $b_u$. Regardless of the distribution of the covariates, the quantity $v_u$ satisfies $v_u \to \gamma$ as $n\to \infty$. We use the fact that $X^tX$ follows the Wishart distribution and satisfies the following properties: 
    \[\mathbb{E}\left[(X^tX)^2\right] = n(n+1)\Sigma^2+n\Sigma\text{tr}(\Sigma) \hspace{2mm};\hspace{2mm}  \mathbb{E}\left[(X^tX)^{-1} \right] = \frac{1}{n-p-1}\Sigma^{-1}.\]
    using the above properties, we can obtain the following limiting expressions of $v_l$ and $b_u$:
    \begin{align*}
        b_u=\frac{\frac{n-1}{n}(p+\frac{n-1}{n})}{n} \text{tr}(\Sigma) \to \gamma c^2 \hspace{2mm};\hspace{2mm}
        v_l= \frac{p}{n-p-1}\to\frac{\gamma}{1-\gamma}.
    \end{align*}
    The limiting expressions $\eta_{\infty}$ and $\dot{\alpha}_{\infty}$ are determined by placing the above limiting values of $v_l$, $v_u$ and $b_u$ in Equations (\ref{Eeta}) and (\ref{E_alpha_dot}) respectively. 
\qed

\subsection{Proof of Theorem~\ref{GenAssym}}\label{appA6}
Let us denote by $\tilde{X}\in \mathbb{R}^{n \times p}$ the random matrix with rows $\{\tilde{x}_i\}_{i=1}^n$, and by $C_{\tilde{X}}$ the matrix of cross terms as in (\ref{CxObs}), with components $\tilde{x}$. Using the fact that $X=\Tilde{X}\Sigma^{1/2}$, we derive the explicit expression of $b_u$ as follows: 
   \begin{align*}
        b_u &=\frac{(n-1)^2}{n^4} \text{tr}\left(   \left[\mathbb{E} (\tilde{X}^t\tilde{X})^2 -  \left( \mathbb{E}[\tilde{X}^t\tilde{X}] \right)^2  \right]  \Sigma  \right)+\frac{1}{n^2} \text{tr}\left(\mathbb{E} (C_{\tilde{X}})^2 \Sigma  \right)\\
        &=\left[\frac{(n-1)^2(p+q-2)+ (n-1)p}{n^3}   \right] \text{tr}\left( \Sigma \right).   
   \end{align*}
    From the above expression we conclude that $b_u\to \gamma c^2$,as $n\to \infty$, under this generating mechanism, as in the Gaussian case. Moreover, it is clear that $v_u\to \gamma$ due to the fact that the quantity $v_u$ is indifferent to the distribution $P_x$. 

    Regarding the quantity $v_l$, we denote the conditional variance component by $v_l(X)$, such that $v_l= \mathbb{E}_X[v_l(X)]$, and write as follows:
    \begin{align*}
        v_l(X) &=\text{tr}\left(\Sigma(X^tX)^{-1}\right) = \text{tr}\left( \Sigma^{\frac{1}{2}}\left(\Tilde{X}^t\Tilde{X}\right)^{-1}\Sigma^{-\frac{1}{2}}\right) = \frac{1}{n}\text{tr}\left(\left(\frac{1}{n}\Tilde{X}^t\Tilde{X}\right)^{-1}\right).
    \end{align*}
     We now note that according to the Marchenko-Pastur theorem \citep{marchenko1967distribution,silverstein1995strong}, under the asymptotic setting, the spectral measure of $\frac{1}{n}\Tilde{X}^t\Tilde{X}$ converges weakly, almost surely, to the Marchenko-Pastur law, depending only on $\gamma$, and bounded away from $0$. Therefore, we can write that as $n, p \to \infty$, almost surely:
     \[v_l(X) = \frac{1}{n}\text{tr}\left(\left(\frac{1}{n}\Tilde{X}^t\Tilde{X}\right)^{-1}\right) \to \gamma s(\gamma),  \]
    where $s(\gamma)$ can be obtained by the Marchenko-Pastur law. In the proof of \cite{hastie2022surprises}, Proposition 2, it was shown that $s(\gamma)=1/(1-\gamma)$, concluding that $v_l(X)$ converges almost surely in $X$ to $\gamma/(1-\gamma)$, which is also the expected value of $v_l(X)$ in the Gaussian case. The almost sure convergence combined with the fact that $v_l(X)\geq 0$ for all possible realizations of $X$ yields that for every $\tilde{\epsilon}>0$, there exists an $n(\tilde{\epsilon})$ such that:
    \[v_l= \mathbb{E}_X[v_l(X)] \geq \frac{\gamma}{1-\gamma} - \tilde{\epsilon} \hspace{2mm},\hspace{2mm} \forall n>n(\tilde{\epsilon}).\]
    Thus, as long as Assumption~\ref{A0n} holds, $v_l$ either converges to a fixed value that is greater than or equal to the one in the Gaussian case, or diverges to large values. Noting that $\eta$ decreases as $v_l$ increases (while $b_u$ and $v_u$ are fixed), we conclude that $\eta$ either converges to a fixed value that is smaller than or equal to $\eta_{\infty}$, or fluctuates between values in the interval $(0,\eta_{\infty}]$ as $n\to \infty$. In the same way, $\dot{\alpha}$ increases as $v_l$ increases, and therefore lies in the interval $[\dot{\alpha}_{\infty},1)$ as $n\to \infty$.
    \qed

\subsection{Proof of Proposition~\ref{ProbConv}}\label{appA7}
For any given training size $n$, the estimator $\hat{\sigma}^2$ is an unbiased estimator of $\sigma^2$, and its total variance over $T=(X,Y)$ is $\mathbb{V}ar(\hat{\sigma}^2)=2\sigma^4/(n(1-\gamma))$. Therefore, $\hat{\sigma}^2$ converges in probability to $\sigma^2$, and also $\hat{\alpha}$ converges in probability to $\dot{\alpha}$. By the definitions of $\dot{\beta}_{\hat{\alpha}(\tau)}$ and $\dot{\beta}_{\dot{\alpha}}$, the difference between the predictions can be written as follows.
\[x_0^t\dot{\beta}_{\dot{\alpha}}  -x_0^t\dot{\beta}_{\hat{\alpha}(\tau)} = (\hat{\alpha}-\dot{\alpha})x_0^t(\hat{\beta}-\breve{\beta}).\]
Under the distributional assumptions taken over $P_x$, the total variance of $x_0^t(\hat{\beta}-\breve{\beta})$ converges to the following fixed value as $n \to \infty$:
\[\mathbb{V}ar \left(x_0^t(\hat{\beta}-\breve{\beta})\right)  \to \frac{\sigma^2\gamma^2}{1-\gamma}  +  \gamma c^2 \tau^2.\]
From the fact that $\hat{\alpha}-\dot{\alpha}$ converges in probability to $0$, and the variance of $x_0^t(\hat{\beta}-\breve{\beta})$ is asymptotically bounded, it is elementary that the product between the two terms also converges in probability to $0$.
\qed

\subsection{Proof of Theorem~\ref{AsymptoticInterpolators}}\label{appA13}
For a given covariance matrix $\Sigma$, let us denote by $U$, the random matrix $U=X\Sigma^{-0.5}$. The properties of the Gaussian distribution imply that the rows of the matrix $U=X\Sigma^{-0.5}$ have a scaled Gaussian distribution with i.i.d. components. Using the connection between $U$ and $X$, we can write the terms $v^I_u$ and $b^I_l$ as follows:
\[v^I_u = \text{tr}\left(\mathbb{E}\left[ (UU^t)^{-1} \right]  \right) \hspace{2mm};\hspace{2mm}  b^I_l =  \text{tr}\left( \mathbb{E}\left[  \Sigma^{1/2} U^t(UU^t)^{-1}U \Sigma^{1/2}\right]  \right) .\]
Taking advantage of the intrinsic properties of the matrix $U$, including its rotational invariance, and the attributes of the generalized inverse of the Wishart matrix (as outlined in \cite{10.1214/11-EJS602}), we can deduce exact expressions for $v^I_u$ and $b^I_l$, as well as their asymptotic values, as described below. 
\begin{align*}
    v^I_u &=\frac{1}{p-n-1}\text{tr}\left(I_n  \right) = \frac{n}{p-n-1}\to \frac{1}{\gamma-1}, \\
    b^I_l &= \frac{\text{tr}\left(\Sigma  \right)}{p}  \text{tr}\left( \mathbb{E}\left[   U^t(UU^t)^{-1}U \right]  \right) = \text{tr}\left(\Sigma  \right) \frac{n}{p}
 \to c^2\frac{1}{\gamma}.     
\end{align*}

For the manner of analyzing the terms $v^I_l$ and $b^I_u$, we adopt the conclusion that arose by \cite{hastie2022surprises} and also by \cite{bartlett2020benign}, that the estimator $\hat{\beta}^I$ is affected by the vanishing eigenvalues of $\Sigma$ in a manner that only the $\tilde{p}$ non-vanishing components of $\Sigma$ are taken into account. Therefore, we can analyze the predictive error of $\hat{\beta}^I$ as is done with $\tilde{p}$-dimensional i.i.d. Gaussian covariates, by taking into account $\tilde{\gamma}$ in place of $\gamma$. More precisely, we can assume that $x\sim MN(\textbf{0}_{\Tilde{p}}, c^2_1(n)I_{\Tilde{p}})$ and the rows of the matrix $U=\frac{1}{c_1(n)}X$ have a scaled Gaussian distribution with i.i.d. components. Therefore, we can write the terms $v^I_l$ and $b^I_u$ as follows:
\begin{align*}
    v^I_l &=\text{tr}\left( \mathbb{E}\left[   U^t(UU^t)^{-2}U \right]  \right) = \frac{1}{\Tilde{p}-n-1}\text{tr}\left(I_n  \right) \to \frac{1}{\tilde{\gamma}-1}, \\
    b^I_u &=  \text{tr}\left(   c^2_1(n)I_{\Tilde{p}}\mathbb{E}_X\left[X^t(XX^t)^{-1}X\right]\right) = 
 c^2_1(n) \cdot n \to c^2\frac{1}{\tilde{\gamma}}.     
\end{align*}
 
For any given combination of $(\sigma^2,\tau^2,c^2,\gamma,\tilde{\gamma})$, we can plug in the above limiting values into the formulas of $\eta^I$ and $\dot{\alpha}_I$ and get the following limiting expressions:
    \begin{align*}
    \eta^I &\to  \eta^I_{\infty} = 1- \frac{(\gamma-\Tilde{\gamma})\sigma^4}{(\gamma-1)^2(\Tilde{\gamma}-1)^2 \left[  \frac{\tau^2c^2}{\gamma \Tilde{\gamma}} +\frac{\sigma^2}{(\gamma-1)(\Tilde{\gamma}-1)} \right]  \left[  \frac{\tau^2c^2(\Tilde{\gamma}-1)}{\Tilde{\gamma}} +\frac{\sigma^2}{\Tilde{\gamma}-1} \right]},\\
    \dot{\alpha}_I &\to \dot{\alpha}_{I,\infty}=\frac{\sigma^2}{(\gamma-1)(\Tilde{\gamma}-1) \left[  \frac{\tau^2c^2}{\gamma \Tilde{\gamma}} +\frac{\sigma^2}{(\gamma-1)(\Tilde{\gamma}-1)} \right]}.
\end{align*}

\qed

\section{Challenging the total information assumption}\label{appB}
In this section we analyze the case where the amount of the unlabeled data, $m$ is comparable to the amount of the labeled data, $n$, and the dimension of the data, $p$, in the specific case of linear models. In this case, the variability in the estimation of parameters like $\mathbb{E}[xx^t]$ is not negligible and affects the predictive error of the semi-supervised estimators. In subsequent analysis, we show that, on the one hand, as anticipated, the predictive error surpasses that observed in the total information scenario. On the other hand, we show that the central argument of Section~\ref{semi-supervised-GLM} holds as long as the quantity of unlabeled data exceeds that of the labeled data. Additionally, for Gaussian covariates specifically, we can precisely express the optimal mixing ratio, and also the ratio between the predictive error of the optimally adjusted semi-supervised estimator and the supervised estimator.

Denoting the set of unlabeled data by $Z\in \mathbb{R}^{m \times p}$, the linear mixed-SSL estimator under discussion can be written as follows:
\begin{align}\label{e7}
\tilde{\beta}_{\alpha}=(1-\alpha)\hat{\beta}+\alpha \tilde{\beta} \hspace{2mm};\hspace{2mm} \tilde{\beta}= \frac{m}{n}\left(Z^tZ \right)^{-1}X^tY.
\end{align}
We note that in the estimator $\tilde{\beta}$ above, we estimate the matrix $H=\mathbb{E}[X^tX]$ by $\frac{n}{m}Z^tZ$, which is now a random matrix and not fixed as in the total information case. For the sake of brevity, we focus on the properties of the estimator $\tilde{\beta}_{\alpha}$ under the random-$\beta$ scenario and assume that the rows of $Z$ are independent of the rows of $X$. In this setting, we can write the predictive error of $\tilde{\beta}_{\alpha}$ as follows:
\begin{align*}
    R(\tilde{\beta}_{\alpha})=r(\tilde{\beta}_{\alpha}) - \frac{1}{2}\tau^2\text{tr}(\Sigma),
\end{align*}
where the reducible error $r(\tilde{\beta}_{\alpha})$ can be written as follows:
\begin{align*}
    r(\tilde{\beta}_{\alpha})=\frac{1}{2}\left[  \alpha^2 \tau^2 b_{\tilde{u}} + \alpha^2 \sigma^2v_{\tilde{u}} +(1-\alpha)^2\sigma^2v_l+ 2\alpha(1-\alpha)v_{\tilde{s}}\right],       
\end{align*}
where:
\begin{align*}
    b_{\tilde{u}}&= \frac{1}{n}\text{tr}\left( H\mathbb{E} \left[ \left(I-\frac{m}{n}(Z^tZ)^{-1}X^tX \right) \left(I-\frac{m}{n}(Z^tZ)^{-1}X^tX \right)^t \right]\right),\\
    v_{\tilde{u}}&=\frac{m^2}{n^3}\text{tr}\left( H\mathbb{E} \left[ (Z^tZ)^{-1}X^tX(Z^tZ)^{-1} \right]\right),\\
    v_{\tilde{s}} &= \frac{m}{n^2}\text{tr}\left(\mathbb{E}\left[ (Z^tZ)^{-1}\right] H\right)
    \end{align*}
and $v_l$ refers to the same quantity as described in Section \ref{semi-supervised-GLM}, and given that $D=I$, it can be expressed as $v_l =\frac{1}{n}\text{tr}\left(\mathbb{E}\left[ (X^tX)^{-1}\right] H\right)$.

When we compute the first and second derivatives of $r(\tilde{\beta}_{\alpha})$ (or $R( \tilde{\beta}_{\alpha})$) with respect to $\alpha$, it results in:
\begin{align*}
    \frac{\partial}{\partial \alpha}r(\tilde{\beta}_{\alpha})  \mid _{\alpha=0} &= -\sigma^2(v_l-v_{\tilde{s}}) , \\
    \frac{\partial}{\partial \alpha}r(\tilde{\beta}_{\alpha})  \mid _{\alpha=1}&= \tau^2 b_{\tilde{u}} +\sigma^2( v_{\tilde{u}}-v_{\tilde{s}} ), \\
    \frac{\partial^2}{\partial \alpha^2}r(\tilde{\beta}_{\alpha})&= \tau^2 b_{\tilde{u}}+\sigma^2(v_l+v_{\tilde{u}}-2v_{\tilde{s}} ).
\end{align*}
We note that the term $v_l+v_{\tilde{u}}-2v_{\tilde{s}}$ is non-negative, since it can be written as an expectation of a trace of the squared form as follows:
\[ v_l+v_{\tilde{u}}-2v_{\tilde{s}} = \mathbb{E}\left[ \text{tr}\left( H^{1/2} Q Q^t H^{1/2}\right) \right],\]
where:
\[Q=\frac{1}{n^{1/2}}(X^tX)^{-1/2}-\frac{m}{n^{3/2}}(Z^tZ)^{-1}(X^tX)^{1/2}.\]
Based on the preceding analysis, it is determined that the second derivative of $r(\tilde{\beta}_{\alpha})$ is positive, which confirms the convexity of $r(\tilde{\beta}_{\alpha})$ with respect to $\alpha$.

We now point to the fact that if the key condition $v_{\tilde{s}}<v_l$ holds, it is always beneficial to slightly move from the supervised estimator, $\hat{\beta}$, towards $\tilde{\beta}$, since the first derivative is negative at $\alpha=0$. If the condition $v_{\tilde{s}}<v_{\tilde{u}}$ holds as well, the unique global minimizer of $r(\tilde{\beta}_{\alpha})$ is necessarily in the open interval $(0,1)$, and admits the following explicit formula:
\begin{align}\label{alphStarFintie}
    \underset{\alpha\in  [0,1]}{\arg \min} \left\{ r(\tilde{\beta}_{\alpha}) \right\}=\frac{\sigma^2(v_l-v_{\tilde{s}}) }{\tau^2 b_{\tilde{u}}+\sigma^2(v_l+v_{\tilde{u}}-2v_{\tilde{s}})}.
\end{align} 
The key condition $v_{\tilde{s}}<v_l$ holds if and only if $\mathbb{E}\left[ (\frac{1}{n}X^tX)^{-1}\right]-\mathbb{E}\left[ (\frac{1}{m}Z^tZ)^{-1}\right]$ is a positive definite matrix, which seems to be a mild requirement in any case of $m>n$. Indeed, we are able to show that this condition holds without any distributional assumption other than Assumption~\ref{A0n}, when $m=Kn$, for $K=2,3,\ldots\ $, as follows:
\begin{align*}
    \mathbb{E}\left[ \left(\frac{1}{n}X^tX\right)^{-1}\right] &= \frac{1}{K}\sum_{k=1}^K \mathbb{E}\left[ \left(\frac{1}{n}X^tX\right)^{-1}\right]\\ &=   \mathbb{E}\left[ \sum_{k=1}^K\frac{1}{K}\left(\frac{1}{n}\sum_{i=s(k)}^{kn}z_iz_i^t\right)^{-1}\right]\\
     & \succeq \mathbb{E}\left[ \left(\sum_{k=1}^K\frac{1}{Kn}\sum_{i=s(k)}^{kn}z_iz_i^t\right)^{-1}\right]= \mathbb{E}\left[ \left(\frac{1}{m}Z^tZ\right)^{-1}\right].
\end{align*}
In the above derivation, $s(k)=1+(k-1)n$, and the symbol $\succeq$ denotes the relationship between positive definite matrices (PD). The second equality is valid due to the i.i.d. mechanism of $X$ and $Z$. The inequality (in the sense of P.D.) is valid because an average of inverse matrices is larger than the inverse of an average of the same matrices, as shown in \cite{nordstrom2011convexity}. Thus, we conclude that the mixed estimator is beneficial in general, without the total information assumption. 

Let us also assume that the distribution $P_x$ satisfies the conditions of Assumption~\ref{AMechanisim_n}, and denote by $\tilde{Z}\in \mathbb{R}^{m \times p}$ the random matrix with scaled i.i.d. entries $\{z_{ij}\}$, such that $Z=\Tilde{Z}\Sigma^{1/2}$. In this case, we can show that the condition $v_{\tilde{s}}<v_{\tilde{u}}$ also holds, as follows:
\begin{align*}
    v_{\tilde{u}}&=\frac{m^2}{n^3}\text{tr}\left( H\mathbb{E} \left[ (Z^tZ)^{-1}X^tX(Z^tZ)^{-1} \right]\right) = \frac{m^2}{n}\text{tr}\left( \mathbb{E} \left[ (\widetilde{Z}^t\widetilde{Z})^{-2} \right]\right) \\
    & \geq \frac{m^2}{n}\text{tr}\left(\left( \mathbb{E} \left[ (\widetilde{Z}^t\widetilde{Z})^{-1} \right]\right)^2\right) \geq  \frac{m^2}{n}\text{tr}\left( \mathbb{E} \left[ (\widetilde{Z}^t\widetilde{Z})^{-1}  \right]  \left( \mathbb{E} \left[ (\widetilde{Z}^t\widetilde{Z}) \right]\right) ^{-1}\right)  \\
    &= \frac{m}{n}\text{tr}\left( \mathbb{E} \left[ (\widetilde{Z}^t\widetilde{Z})^{-1} \right]\right) = \frac{m}{n}\text{tr}\left( \mathbb{E} \left[ (Z^tZ)^{-1} \right] \Sigma\right) =  v_{\tilde{s}}.
\end{align*}
From the above, we conclude that the theoretical results regarding the advantages of using the unlabeled data are valid for a wide class of distributions, without the requirement of the total information assumption.

In the more specific case of Gaussian covariates, we can explicitly write all relevant terms: $b_{\tilde{u}}$, $v_{\tilde{u}}$, $v_{\tilde{s}}$, and $v_l$. Therefore, we can get more insights on how the mixed-SSL estimator is affected by the size of the unlabeled data. In particular, we can compare between $v_l$ and $v_{\tilde{s}}$ as follows:
\[v_l-v_{\tilde{s}} = p\left(\frac{1}{n-p-1} - \frac{1}{n-p\frac{m}{n}-1} \right),\]
which is positive as long as $m>n$, revealing that the mixed-SSL is beneficial when the amount of unlabeled data exceeds that of the labeled data. In order to obtain a succinct view on the properties of $\tilde{\beta}_{\alpha}$, we consider the asymptotic setting in which $p/n \to \gamma\in(0,1)$ and $p/m \to \gamma' < \gamma$ as $n,m,p\to \infty$, and assumption~\ref{A01} is satisfied. In this context, we are able to show that:
\begin{align*}
    b_{\tilde{u}} &\to c^2 \frac{1+(1-\gamma')^3-2(1-\gamma')^2+\gamma}{(1-\gamma')^3},\\
    v_{\tilde{u}}&\to\frac{\gamma}{(1-\gamma')^3} \hspace{2mm}; \hspace{2mm}
     v_{\tilde{s}} \to\frac{\gamma}{1-\gamma'}.
\end{align*}
From the above, it is clear that for any mixing ratio $\alpha$, the value of $\tilde{r}(\alpha)$ progressively increases with $\gamma'$ and matches $\dot{r}(\alpha)$ when $\gamma'=0$. Similarly, $\arg\min \left\{ r(\tilde{\beta}_{\alpha}) \right\}$ aligns with $\dot{\alpha}$ from the total information scenario as $\gamma'$ approaches $0$. Moreover, these limiting values are useful for defining the limiting values of $\arg\min \left\{ r(\tilde{\beta}_{\alpha}) \right\}$ in any specific configuration of $(\gamma,\gamma')$, as outlined in Equation (\ref{alphStarFintie}).

\section{Quadratic Approximation Derivations}\label{appC}
Let us denote $L(\beta_T,x_0)=G(x_0^t\beta_T) - x_0^t\beta_T g(x_0^t\beta)$, the random error of an estimator $\beta_T$, and a covariate vector $x_0$, such that $R_{\beta}(\beta_T)= \mathbb{E}_{x_0,T}\left[ L(\beta_T,x_0) \right ]$. The predictive error of an estimator $\beta_T$ can be approximated using a quadratic expansion around the true vector $\beta$, as follows:
\[R_{\beta}(\beta_T) \approx R_{\beta}(\beta)+ \mathbb{E}\left[(\beta_T-\beta)^tL'(\beta,x_0)\right]    +\frac{1}{2} \mathbb{E}\left[(\beta_T-\beta)^t L''(\beta,x_0)(\beta_T-\beta)\right],\]
where the derivatives are taken with respect to $\beta_T$. Noting that $L'(\beta,x_0)= 0$, we obtain the following approximate expression for an estimator $\beta_T$:
\begin{align} \nonumber
    R_{\beta}(\beta_T) 
    \approx  & R_{\beta}(\beta)+\frac{1}{2} \mathbb{E}\left[(\beta_T-\beta)^tL''(\beta,x_0)(\beta_T-\beta)\right] \\  \nonumber \approx & R_{\beta}(\beta) +\frac{1}{2}\mathbb{E}_{T}\left[ (\beta_T-\beta)^t E_{x_0}\left[g'(x_0^t\beta)x_0x_0^t \right]  (\beta_T-\beta) \right] \\  \nonumber 
    =& R_{\beta}(\beta) + \frac{1}{2n}\mathbb{E}_{T}\left[ (\beta_T-\beta)^tH(\beta_T-\beta)\right] \\ \label{V_bet_approx} =& R_{\beta}(\beta) +  \underbrace{  \frac{1} {2n}\text{tr}\left( H  \mathbb{E}_{X}\left[ \mathbb{V}ar\left( \beta_T \mid X\right)\right] \right)
  }_\text{$V_{\beta}(\beta_T)$} \\ \label{B_bet_approx}
    &  + \underbrace{ \frac{1} {2n}\text{tr}\left( H \mathbb{E}_{X}\left[ \left(\mathbb{E}[\beta_T \mid X] - \beta   \right)\left(\mathbb{E}[\beta_T \mid X] - \beta   \right)^t \right]  \right) }_\text{$B_{\beta}(\beta_T)$}.
\end{align}

The quadratic approximation of $\hat{L}(\hat{\beta})$, defined in Equation (\ref{L_hat_Def}), around the real $\beta$, can be written as follows:  
\begin{equation*}
    \hat{L}(\hat{\beta}) \approx L(\beta)+(\hat{\beta}-\beta)^tL'(\beta) +\frac{1}{2}(\hat{\beta}-\beta)^tL''(\beta)(\hat{\beta}-\beta).
\end{equation*}
Differentiating  both sides according to $\hat{\beta}$, since $\hat{L}'(\hat{\beta})=\hat{S}(\hat{\beta}) =\textbf{0}$, we get:
\begin{align*}
    \textbf{0} &\approx \hat{L}'(\beta)- \hat{L}''(\beta)\beta+ \hat{L}''(\beta)\hat{\beta} \implies\\
    \hat{\beta} &\approx  \beta - \left(\hat{L}''(\beta)\right)^{-1}\left[\hat{L}'(\beta) \right] 
    = \beta - \left(X^tDX\right)^{-1}X^t(\mu-Y):= \beta -\hat{a}.
    \end{align*}
In the same manner, we can apply quadratic approximation to $\breve{L}(\beta)$ and $L^{M}_{\alpha}(\beta)$, defined in Equations (\ref{L_breve_Def}) and (\ref{L_Mixed_Def}) respectively, and show  that:
\begin{align*}
    \Breve{\beta} & \approx  \beta - \left(\breve{L}''(\beta)\right)^{-1}\left[\breve{L}'(\beta) \right] \\&=\beta - H^{-1}\left(\mathbb{E}_X[X^t\mu] -n \widehat{\mathbb{C}ov}(X,Y) \right):=  \beta - H^{-1}\zeta_Y := \beta -\Breve{a},\\
    \dot{\beta}_{\alpha} & \approx \beta - \alpha \breve{a} -(1-\alpha)\hat{a} :=   \beta - \dot{a}_{\alpha}
    \\  \Ddot{\beta}_{\alpha} & \approx  \beta - \left( \left(L^{M}_{\alpha} \right)
 ''(\beta)\right)^{-1}\left [ \left(L^{M}_{\alpha} \right)'(\beta) \right] 
 \\ &= \beta - \left(\alpha \breve{L}''(\beta) +(1-\alpha) \hat{L}''(\beta) \right)^{-1}\left[ \alpha \breve{L}'(\beta) +(1-\alpha) \hat{L}'(\beta)  \right]
 \\ &=  \beta - S_{\alpha} \left ( \alpha(\mathbb{E}_X[X^t\mu]+n\overline{X}\cdot \overline{Y})  + (1-\alpha)X^t\mu- X^tY  \right) \\ &= \beta - S_{\alpha}\left( \alpha\zeta_Y+(1-\alpha)X^t(\mu-Y)\right) := \beta -\Ddot{a}_{\alpha}.
\end{align*}

 We note that in all derivations of the Taylor expansion above, the quadratic term depends on $\beta$ through $g'$ and the cubic term necessarily depends on $\beta$ through $g''$.  Therefore, when the link function meets the condition where the magnitude of $g''$ is both limited and insignificant compared to $g'$, employing the quadratic approximation is justified for obtaining theoretical insights and for comparing the predictive performance of various estimators.

By substituting $\beta_T$ in (\ref{V_bet_approx}) with the approximate expressions of $\dot{\beta} _{\alpha}$, and $\Ddot{\beta} _{\alpha}$, we can write the following approximate variance terms: 
\begin{align*}
     V_{\beta}(\dot{\beta}_{\alpha})\approx  &\frac{1} {2n}\text{tr}\left( H  \mathbb{E}_{X}\left[ \mathbb{V}ar\left(\dot{a}_{\alpha} \mid X\right)\right] \right) \\ = &\frac{1} {2n}\text{tr}\left( H  \mathbb{E}_{X}\left[ \mathbb{V}ar\left(\alpha\breve{a}+(1-\alpha)\hat{a} \mid X\right)  \right] \right) \\ =& \hspace{1mm} \frac{1} {2n}\alpha^2\text{tr}\left(H^{-1}\mathbb{E}_X\left[ \mathbb{V}ar(\zeta_y \mid X)\right] \right)  \\&+\frac{1} {2n}(1-\alpha)^2\text{tr}\left(H\mathbb{E}_X\left[ \left(X^tDX\right)^{-1}X^t \mathbb{V}ar(Y \mid X)X\left(X^tDX\right)^{-1}\right] \right)  \\&+\frac{1} {n}\alpha(1-\alpha)\text{tr}\left(\mathbb{E}_X\left[ \left(X^tDX\right)^{-1}X^t \mathbb{C}ov(Y,\zeta_y \mid X)\right] \right),
    \\ V_{\beta}(\ddot{\beta}_{\alpha}) \approx  & \hspace{1mm}\frac{1} {2n}\text{tr}\left( H  \mathbb{E}_{X}\left[ \mathbb{V}ar\left(\ddot{a}_{\alpha} \mid X\right)\right] \right) \\ = & \hspace{1mm} \frac{1} {2n}\alpha^2\text{tr}\left(\mathbb{E}_X\left[  S_{\alpha}HS_{\alpha}\mathbb{V}ar(\zeta_y \mid X)\right] \right)  \\&+\frac{1} {2n}(1-\alpha)^2\text{tr}\left(H\mathbb{E}_X\left[ S_{\alpha}X^t \mathbb{V}ar(y \mid X)XS_{\alpha}\right] \right)  \\&+\frac{1}{n}\alpha(1-\alpha)\text{tr}\left(\mathbb{E}_X\left[ S_{\alpha}HS_{\alpha}X^t\mathbb{C}ov(Y,\zeta_y \mid X)\right] \right).
\end{align*}
By using the fact that $\mathbb{V}ar(Y \mid X)=\sigma^2D$, we can obtain the following:
\begin{align*}
     V_{\beta}(\dot{\beta}_{\alpha})\approx  &  \hspace{1mm} \frac{1}{2n}\alpha^2 \sigma^2 \frac{n-1}{n} \text{tr}\left(H^{-1}\mathbb{E}_X\left[ X^tDX\right] \right)\\&+\frac{1}{2n}(1-\alpha)^2\sigma^2\text{tr}\left(H\mathbb{E}_X\left[ \left(X^tDX\right)^{-1}\right] \right)  \\&+\frac{1}{n}\alpha(1-\alpha)\sigma^2 \frac{n-1}{n}\text{tr}\left(\mathbb{E}_X\left[ \left(X^tDX\right)^{-1}X^tDX\right] \right)   \\
     =& \hspace{1mm} \frac{1}{2}\sigma^2\left[\alpha^2 v_ u +(1-\alpha)^2v_l +2\alpha(1-\alpha)v_u\right] = \dot{V}_{\beta}(\alpha)
    \\ V_{\beta}(\ddot{\beta}_{\alpha}) \approx  & \hspace{1mm}\frac{1}{2n}\alpha^2 \sigma^2 \frac{n-1}{n} \text{tr}\left(\mathbb{E}_X\left[ S_{\alpha}HS_{\alpha} X^tDX\right] \right)\\&+\frac{1}{2n}(1-\alpha)^2\sigma^2\text{tr}\left(\mathbb{E}_X\left[ S_{\alpha}HS_{\alpha} X^tDX\right] \right)  \\&+\frac{1}{n}\alpha(1-\alpha)\sigma^2 \frac{n-1}{n}\text{tr}\left(\mathbb{E}_X\left[ S_{\alpha}HS_{\alpha} X^tDX\right] \right) \\
    =&  \hspace{1mm} \frac{\sigma^2}{2n}
    \left[1-2\alpha+\alpha^2+\frac{n-1}{n}(\alpha^2+2\alpha-2\alpha^2) \right] \text{tr}\left(\mathbb{E}_X\left[ S_{\alpha}HS_{\alpha} X^tDX\right] \right) \\
    =& \hspace{1mm} \frac{ \xi_{\alpha}\sigma^2}{2n}\text{tr}\left(\mathbb{E}_X\left[ S_{\alpha}HS_{\alpha} X^tDX\right] \right)=\ddot{V}_{\beta}(\alpha). 
\end{align*}

By replacing $\beta_T$ in (\ref{B_bet_approx}) with the approximated expressions for $\dot{\beta} _{\alpha}$ and $\Ddot{\beta} _{\alpha}$, and employing the property $\mathbb{E}[Y \mid X]=\mu$, we derive the following form of the approximate bias terms: 
\begin{align*}
     B_{\beta}(\dot{\beta}_{\alpha}) \approx  & \hspace{1mm} \frac{1} {2n}\text{tr}\left( H  \mathbb{E}_{X}\left[ \mathbb{E}\left[\dot{a}_{\alpha} \mid X\right]\mathbb{E}\left[\dot{a}^t_{\alpha} \mid X\right]\right] \right) \\=& \hspace{1mm} \frac{\alpha^2} {2n}\text{tr}\left( H  \mathbb{E}_{X}\left[ \mathbb{E}\left[\breve{a} \mid X\right]\mathbb{E}\left[\breve{a}^t \mid X\right]\right] \right) \\= & \hspace{1mm}\frac{\alpha^2} {2n}\text{tr}\left(H^{-1}\mathbb{E}_X\left[ \mathbb{E}[\zeta_Y\zeta_Y^t \mid X]\right] \right) \\ =& \hspace{1mm} \frac{\alpha^2} {2n}\text{tr}\left(H^{-1}\mathbb{E}_X\left[\zeta\zeta^t\right] \right) = \dot{B}_{\beta}(\alpha),     
    \\ B_{\beta}(\ddot{\beta}_{\alpha}) \approx  & \hspace{1mm} \frac{1}{2n}\text{tr}\left( H  \mathbb{E}_{X}\left[ \mathbb{E}\left[\ddot{a}_{\alpha} \mid X\right]\mathbb{E}\left[\ddot{a}^t_{\alpha} \mid X\right]\right] \right) \\=& \hspace{1mm}  \frac{\alpha^2} {2n}\text{tr}\left(H\mathbb{E}_X\left[ S_{\alpha}\mathbb{E}[\zeta_Y\zeta_Y^t \mid X]S_{\alpha}\right] \right) \\ =& \hspace{1mm}  \frac{\alpha^2}{2n}\text{tr}\left(\mathbb{E}_{X}\left[ S_{\alpha}  H S_{\alpha} \zeta \zeta^t \right]\right) = \ddot{B}_{\beta}(\alpha).
\end{align*}

In the context of estimating the noise parameter $\sigma^2$ , we derive the following quadratic approximation of $RSS(\hat{\beta})$ around the real $\beta$:
\begin{align*}
    RSS(\hat{\beta}) & = \mid  \mid g(X\hat{\beta}) -Y  \mid  \mid _2^2 \\ &\approx  \mid  \mid \mu -Y  \mid  \mid _2^2 + 2 (\mu -Y )^tDX(\hat{\beta}-\beta)+(\hat{\beta}-\beta)^tX^tD^2X(\hat{\beta}-\beta),
\end{align*}
and substitute $\hat{\beta}-\beta=(X^tDX)^{-1}X^t(Y-\mu)$ to get the following approximate expression:
\begin{align*}
RSS(\hat{\beta})\approx\text{tr}\Bigg( &(\mu-Y)(\mu-Y)^t\Big(I-2DX(X^tDX)^{-1}X^t \\&+  X(X^tDX)^{-1}X^tD^2X(X^tDX)^{-1}X^t\Big)  \Bigg).
\end{align*}
Now, using the fact that $\mathbb{E}\left[Y \mid X\right]=\mu$ and also that $\mathbb{E}[YY^t \mid X]=\mu \mu^t +D\sigma^2$, we can obtain the following approximation:
\[\mathbb{E}\left[RSS(\hat{\beta}) \mid X\right] \approx \sigma^2\left[\text{tr}\left( D  \right)-\text{tr}\left(X^tD^2X(X^tDX)^{-1}   \right) \right]. \]

\end{appendix}

\begin{acks}[Acknowledgments]
The authors would like to thank the anonymous referees, an Associate
Editor and the Editor for their constructive comments that improved the
quality of this paper.
\end{acks}

\begin{funding}
The authors were supported by \textit{the Israeli Science Foundation grant 3250/24} and \textit{the Israel Council for Higher Education Data-Science Centers}.
\end{funding}

\begin{supplement}
\stitle{Codes for Figures}
\sdescription{The file \say{CodesFor\_MixedSSL\_Paper.ipynb} contains all the Python codes used to generate the results shown in this paper's figures. Access to CelebA dataset is required in order to generate Figure \ref{CelebAFig}. The dataset for Figures \ref{f6} and \ref{f7} is included as an additional supplement.}
\end{supplement}

\begin{supplement}
\stitle{Customized Netflix dataset}
\sdescription{The file \say{Netflix\_X\_Vote.csv} contains the ratings of $12,931$ users that rated the movie \say{Miss Congeniality}, for the $184$ movies with the minimal number of missing values among the full Netflix dataset. The file \say{Netflix\_y.csv} contains the ratings of the movie \say{Miss Congeniality} for the same users.}
\end{supplement}

\bibliographystyle{imsart-nameyear} 
\bibliography{reference}       

\end{document}